\documentclass{article}

\usepackage{arxiv}
\usepackage[utf8]{inputenc} 
\usepackage[T1]{fontenc}    
\usepackage{hyperref}       
\usepackage{url}            
\usepackage{booktabs}       
\usepackage{amsmath,amssymb,amsfonts} 
\usepackage{nicefrac}       
\usepackage{microtype}      
\usepackage{lipsum}
\usepackage{authblk}
\usepackage{graphicx}
\usepackage{comment}
\usepackage{xcolor}         
\usepackage{adjustbox}
\usepackage{subcaption}
\usepackage{float}
\usepackage{algorithmic}
\usepackage[compatibility=false]{caption} 
\usepackage{cite}
\usepackage{textcomp}
\usepackage{siunitx}
\usepackage{balance}
\usepackage{tikz}
\usetikzlibrary{calc,quotes,angles,positioning} 
\usepackage{multirow}
\usepackage{makecell}
\usepackage{array}
\usepackage{mathtools}
\usepackage{acronym}
\usepackage{type1cm}
\usepackage{url}
\acrodef{ADD}{average decision-directed}
\acrodef{ADD-TT}{average decision-directed with time truncation}
\acrodef{Adam}{adaptive moment estimation}
\acrodef{AE}{auto-encoder}
\acrodef{AE-DNN}{auto-encoder deep neural network}
\acrodef{AGC}{automatic gain control}
\acrodef{AI}{artificial intelligence}
\acrodef{ALS}{accurate least squares}
\acrodef{ARM}{advanced RISC machine}
\acrodef{AoI}{age of information}
\acrodef{AWGN}{additive white Gaussian noise}
\acrodef{BER}{bit error rate}
\acrodef{BO}{Bayesian optimisation}
\acrodef{BPSK}{binary phase shift keying}
\acrodef{BSM}{basic safety message}
\acrodef{CAIR}{Centre for Artificial Intelligence Research}
\acrodef{CAM}{class activation mapping}
\acrodef{CDF}{cumulative distribution function}
\acrodef{CDP}{constructed data pilots}
\acrodef{C-ITS}{cooperative intelligent transport systems}
\acrodef{C-V2X}{cellular vehicle-to-everything}
\acrodef{CCH}{control channel}
\acrodef{CEPT}{Conference of Postal and Telecommunications Administrations}
\acrodef{CFO}{carrier frequency offset}
\acrodef{ChannelNet}{channel network}
\acrodef{CIR}{channel impulse response}
\acrodef{CFR}{channel frequency response}
\acrodef{CNN}{convolutional neural network}
\acrodef{CoE}{Centre of Excellence}
\acrodef{CP}{cyclic prefix}
\acrodef{CPU}{central processing unit}
\acrodef{CSI}{channel state information}
\acrodef{DENM}{decentralised environmental notification message}
\acrodef{DeepLIFT}{deep learning important features}
\acrodef{DFT}{discrete Fourier transform}
\acrodef{DNN}{deep neural network}
\acrodef{DNN-STA}{deep neural network spectral temporal averaging}
\acrodef{DL}{deep learning}
\acrodef{DN-CNN}{denoising convolutional neural network}
\acrodef{DPA}{data-pilot aided}
\acrodef{DPA-TCN}{data-pilot-aided temporal convolutional network}
\acrodef{DSP}{digital signal processor}
\acrodef{DSRC}{dedicated short-range communications}
\acrodef{EDCA}{enhanced distributed channel access}
\acrodef{ETSI}{European Telecommunications Standards Institute}
\acrodef{FBF}{frame-by-frame}
\acrodef{FCC}{Federal Communications Commission}
\acrodef{FEC}{forward error correction}
\acrodef{FFT}{fast Fourier transform}
\acrodef{FLOPs}{floating-point operations}
\acrodef{FNN}{feedforward neural network}
\acrodef{FPGA}{field-programmable gate array}
\acrodef{GPU}{graphics processing unit}
\acrodef{GradCAM}{gradient-weighted class activation mapping}
\acrodef{GRACE}{gradient-based channel estimation}
\acrodef{GRU}{gated recurrent unit}
\acrodef{HLS}{high-level synthesis}
\acrodef{ICI}{inter-carrier interference}
\acrodef{iCDP}{improved constructed data pilots}
\acrodef{IDFT}{inverse discrete Fourier transform}
\acrodef{IEEE}{Institute of Electrical and Electronics Engineers}
\acrodef{IFFT}{inverse fast Fourier transform}
\acrodef{ISI}{inter-symbol interference}
\acrodef{ID}{in-distribution}
\acrodef{ITS}{intelligent transportation systems}
\acrodef{LFSR}{linear feedback shift register}
\acrodef{LIME}{local interpretable model-agnostic explanations}
\acrodef{LTV}{linear time-variant}
\acrodef{LMMSE}{linear minimum mean-square error}
\acrodef{LOS}{line-of-sight}
\acrodef{LR}{learning rate}
\acrodef{LRP}{layer-wise relevance propagation}
\acrodef{LS}{least squares}
\acrodef{LSTM}{long short-term memory}
\acrodef{LTS}{long training symbol}
\acrodef{MAE}{mean absolute error}
\acrodef{MAC}{medium access control}
\acrodef{MCS}{modulation and coding scheme}
\acrodef{MIMO}{multiple-input multiple-output}
\acrodef{SISO}{single-input single-output}
\acrodef{ML}{maximum likelihood}
\acrodef{MLP}{multilayer perceptron}
\acrodef{NHTSA}{National Highway Traffic Safety Administration}
\acrodef{MMSE}{minimum mean-square error}
\acrodef{MMSE-VP}{minimum mean-square error using virtual pilots}
\acrodef{MSE}{mean-square error}
\acrodef{MPDU}{MAC protocol data unit}
\acrodef{NN}{neural network}
\acrodef{NLOS}{non-line-of-sight}
\acrodef{NITheCS}{National Institute for Theoretical and Computational Sciences}
\acrodef{NMSE}{normalised mean-square error}
\acrodef{NSGA-II}{non-dominated sorting genetic algorithm II}
\acrodef{NWU}{North-West University}
\acrodef{OOD}{out-of-distribution}
\acrodef{OFDM}{orthogonal frequency-division multiplexing}
\acrodef{PAPR}{peak-to-average power ratio}
\acrodef{PCA}{principal component analysis}
\acrodef{PDF}{probability density function}
\acrodef{PDP}{power delay profile}
\acrodef{PER}{packet error rate}
\acrodef{PHY}{physical layer}
\acrodef{PPDU}{physical layer protocol data unit}
\acrodef{PRR}{packet reception ratio}
\acrodef{PSD}{power spectral density}
\acrodef{PCF}{power control first}
\acrodef{PSNR}{peak signal-to-noise ratio}
\acrodef{QAM}{quadrature amplitude modulation}
\acrodef{QoS}{quality of service}
\acrodef{QPSK}{quadrature phase shift keying}
\acrodef{R2V}{roadside-to-vehicle}
\acrodef{RB}{radial basis function}
\acrodef{ReLU}{rectified linear unit}
\acrodef{RMSE}{root mean square error}
\acrodef{RMS-DS}{root mean-square delay spread}
\acrodef{RNN}{recurrent neural network}
\acrodef{RS}{reliable subcarriers}
\acrodef{RSU}{roadside unit}
\acrodef{RTV}{roadside-to-vehicle}
\acrodef{RTV-EX}{roadside-to-vehicle expressway}
\acrodef{RTV-SUS}{roadside-to-vehicle suburban street}
\acrodef{RTV-UC}{roadside-to-vehicle urban canyon}
\acrodef{REACH}{relevance explainability analysis for channel estimation}
\acrodef{SBS}{symbol-by-symbol}
\acrodef{SCH}{service channel}
\acrodef{SF}{SIGNAL field}
\acrodef{SFO}{sampling frequency offset}
\acrodef{SGD}{stochastic gradient descent}
\acrodef{SHAP}{Shapley additive explanations}
\acrodef{SLS}{simple least squares}
\acrodef{SNR}{signal-to-noise ratio}
\acrodef{SoC}{system on chip}
\acrodef{SOTA}{state-of-the-art}
\acrodef{SR-CNN}{super-resolution convolutional neural network}
\acrodef{SR-ConvLSTM}{super-resolution convolutional long short-term memory}
\acrodef{STA}{spectral temporal averaging}
\acrodef{STA-MLP}{spectral temporal averaging multilayer perceptron}
\acrodef{STS}{short training symbol}
\acrodef{TA}{time averaging}
\acrodef{TA-TDFT}{temporal averaging truncated discrete Fourier transform}
\acrodef{TCH}{control channel}
\acrodef{TCN}{temporal convolutional network}
\acrodef{T-DFT}{truncated discrete Fourier transform}
\acrodef{TDL}{tapped delay line}
\acrodef{TDR}{transmission data rate}
\acrodef{TPE}{tree-structured Parzen estimator}
\acrodef{TRFI}{time-domain reliable frequency-domain interpolation}
\acrodef{TRFI-DNN}{time-domain reliable frequency-domain interpolation DNN}
\acrodef{US}{uncorrelated scattering}
\acrodef{UTC}{Coordinated Universal Time}
\acrodef{URT}{unreliable subcarriers}
\acrodef{V2I}{vehicle-to-infrastructure}
\acrodef{V2P}{vehicle-to-pedestrian}
\acrodef{V2N}{vehicle-to-network}
\acrodef{V2V}{vehicle-to-vehicle}
\acrodef{VTV}{vehicle-to-vehicle}
\acrodef{V2X}{vehicle-to-everything}
\acrodef{VTV-EX}{vehicle-to-vehicle expressway}
\acrodef{VTV-SDWW}{vehicle-to-vehicle same direction with wall}
\acrodef{VTV-UC}{vehicle-to-vehicle urban canyon}
\acrodef{WandB}{Weights and Biases}
\acrodef{WAVE}{wireless access in vehicular environments}
\acrodef{WI}{weighted interpolation}
\acrodef{WSS}{wide-sense stationary}
\acrodef{WSSUS}{wide-sense stationary uncorrelated scattering}
\acrodef{XAI}{explainable artificial intelligence}
\acrodef{XAI-CHEST}{explainable AI for channel estimation}

\title{REACH: Interpretability-Driven Feature Identification and Architecture Compression for Multi-Channel Vehicular Channel Estimation}

\author[1,2,3]{\textbf{S.~A.~Ngorima}}
\author[1]{\textbf{A.~S.~J.~Helberg}}
\author[1,2,3]{\textbf{M.~H.~Davel}}

\affil[1]{Faculty of Engineering, North-West University, South Africa}
\affil[2]{Centre for Artificial Intelligence Research, South Africa}
\affil[3]{National Institute for Theoretical and Computational Sciences, South Africa}
\affil[ ]{\texttt{aldringorima@gmail.com}}

\date{~} 

\begin{document}
\maketitle 
\begingroup
\renewcommand{\thefootnote}{}
\footnote{This work has been submitted to the IEEE for possible publication. Copyright may be transferred without notice, after which this version may no longer be accessible.}
\endgroup

\begin{abstract}
Multi-channel mixed-SNR training improves out-of-distribution (OOD) generalisation of deep learning channel estimators for IEEE 802.11p vehicular communications~\cite{ngorima_access2026}, yet the internal mechanism responsible for this remains unexplained. This work presents REACH (Relevance-based Explanation and Architectural Compression for cHannel estimators), a gradient-based interpretability framework that operates at two levels. Input-level attribution identifies a subset of time--frequency features consistently relevant across all evaluated channel conditions, enabling input dimensionality reduction with minimal performance loss. Filter-level attribution reveals a near-universal internal representation, providing a representational account of the observed OOD generalisation.
Guided by the resulting filter taxonomy, relevance-guided architecture compression substantially reduces both the number of parameters and the number of floating-point operations (FLOPs) with sub-1 dB normalised mean square error (NMSE) degradation, and OOD generalisation degrades more slowly than within-distribution accuracy under increasing compression.
\end{abstract}

\section{INTRODUCTION}
\label{sec:intro}

Vehicular communication systems face a fundamental tension between estimation accuracy and computational efficiency. In high-mobility environments governed by standards such as IEEE 802.11p and emerging \ac{V2X} specifications, \ac{CSI} must be estimated rapidly and accurately to enable reliable data transmission. Severe Doppler spreads reaching several kHz, multipath propagation, and strict latency constraints together render classical estimators inadequate~\cite{pan2021channel}. Linear interpolation fails under fast fading, while \ac{MMSE} methods incur prohibitive complexity for real-time deployment~\cite{gizzini2020deep}.

\Ac{DL} has emerged as a compelling alternative, with \ac{NN}-based channel estimators demonstrating significant gains over classical approaches in vehicular scenarios~\cite{soltani2019deep,ye2018power}. Despite this progress, the black-box nature of \ac{DL} estimators obscures which features drive performance, preventing principled complexity reduction for resource-constrained vehicular hardware. Interpretability addresses this obstacle directly: by ranking input features and internal components according to their contribution to the network output, attribution methods provide a relevance ordering that can guide which features and filters to retain and which to remove, replacing magnitude heuristics and trial-and-error compression with reductions justified by the model's own evidence of what it relies on. This paper addresses this challenge directly. We present REACH (Relevance-based Explanation and Architectural Compression for cHannel estimators), a combined interpretability and architecture compression framework applied to a DPA-RDCNN model trained across multiple channel models~\cite{ngorima_access2026}. The DPA-RDCNN, introduced in~\cite{ngorima_access2026}, couples \ac{DPA} refinement of an initial pilot-based channel estimate with a residual dilated temporal convolutional backbone that processes complete \ac{OFDM} frames, providing deterministic inference latency suitable for real-time vehicular modems. REACH operates at two levels: input feature attribution, which identifies which time--frequency features the model relies on across channel conditions, and internal filter attribution, which reveals the representational structure responsible for \ac{OOD} generalisation.
%

\subsection*{CONTRIBUTIONS}
This work introduces REACH, a gradient-based interpretability and architecture compression framework for multi-channel-trained deep learning channel estimators. Within this framework, the paper makes three contributions:
\begin{enumerate}
    \item \textbf{Input-level interpretability and dimensionality reduction.}
    An attribution procedure that identifies cross-channel important input features consistently relevant across six vehicular channel models, enabling substantial input dimensionality reduction with minimal performance degradation.

    \item \textbf{Filter-level interpretability.}
    An intermediate-layer attribution procedure that classifies the internal convolutional filters of a multi-channel-trained DPA-RDCNN into universal, environment-specific, and redundant categories, characterising the representational structure that supports cross-channel generalisation under multi-channel training.

    \item \textbf{Relevance-guided architecture compression with preserved cross-channel behaviour.} A compact model family whose target filter width is set by the redundant filter fraction identified by the taxonomy, achieving up to 60.2\% parameter reduction and 60.4\% FLOP reduction with sub-1 dB \ac{NMSE} degradation, while the ID--OOD gap remains small at every compression level.
\end{enumerate}

\subsection*{PAPER ORGANISATION}
Section~\ref{sec:background} reviews related work on channel estimation, model interpretability, and structured pruning. Section~\ref{sec:system_model} establishes the system model and the channel estimator that REACH analyses. Section~\ref{sec:reach} develops the REACH attribution methodology at both the input and the filter levels. Section~\ref{sec:results} reports attribution results, both input-level and filter-level. Section~\ref{sec:pruning} applies the resulting filter taxonomy to architecture compression. Section~\ref{sec:discussion} discusses findings and Section~\ref{sec:conclusion} concludes.

\section{BACKGROUND AND RELATED WORK}
\label{sec:background}

\subsection{DL-BASED CHANNEL ESTIMATION FOR IEEE~802.11p}

\ac{DL}-based estimators for IEEE 802.11p were initially developed using \acp{FNN} applied to individual \ac{OFDM} symbols. Ye \emph{et al.}~\cite{ye2018power} introduced \acp{FNN} for \ac{OFDM} channel estimation under non-linear distortion, while Soltani \emph{et al.}~\cite{soltani2019deep} treated the time--frequency grid as an image and applied \acp{CNN} to refine classical pilot-based estimates. Gizzini \emph{et al.}~\cite{gizzini2020deep} extended this line of work to IEEE 802.11p specifically, reporting \ac{NMSE} and \ac{BER} gains over classical \ac{LS}, \ac{DPA}, \ac{STA}, and \ac{TRFI} methods.

For high-mobility vehicular channels, recurrent architectures have been widely applied due to their capacity to model temporal dynamics. Gizzini \emph{et al.}~\cite{gizzini2021temporal} combined \ac{LSTM} networks with temporal averaging to track rapid channel variations, and Hou \emph{et al.}~\cite{9810508} proposed a \ac{GRU}-based estimator exploiting the gating mechanism to selectively retain relevant temporal information. A subsequent study~\cite{10314524} compared \ac{LSTM}, \ac{GRU}, and bidirectional variants across multiple channel models, establishing recurrent estimators as a strong benchmark for doubly selective vehicular environments. Pan \emph{et al.}~\cite{pan2021channel} further investigated joint \ac{LSTM}--\ac{FNN} designs for multi-channel scenarios.

More recently, \ac{TCN}-based estimators have emerged as a parallelisable alternative to recurrent architectures. Ngorima \emph{et al.}~\cite{Ngorima2024} applied \acp{TCN} to \ac{V2X} channel estimation, demonstrating competitive accuracy with reduced latency relative to \ac{LSTM} baselines. The DPA-RDCNN architecture~\cite{ngorima_access2026} builds on this direction by combining \ac{DPA} channel estimation with a stack of residual dilated 1D convolutions that refine the \ac{DPA} estimate, and by training jointly across multiple channel models.

Across the architecture families surveyed above, the dominant evaluation practice remains single-channel training and testing~\cite{soltani2019deep, gizzini2020deep, gizzini2021temporal, 9810508}, leaving generalisation under distribution shift largely uncharacterised. Multi-channel mixed-\ac{SNR} training has been shown to improve \ac{OOD} generalisation across diverse vehicular propagation conditions~\cite{ngorima_access2026}, yet the internal mechanism responsible for this transfer remains unexplained. Furthermore, across all architecture families surveyed above, the absence of principled interpretability prevents systematic identification of which features and internal components drive estimation accuracy.

\subsection{INTERPRETABILITY FOR DL CHANNEL ESTIMATION}
\Ac{XAI} methods are generally categorised into model-agnostic and model-specific approaches. Model-agnostic techniques, such as perturbation-based schemes, treat the underlying model as a black box and analyse input-output relationships by inducing noise or masking features~\cite{adadi2018peeking}. In the context of the physical layer, \ac{XAI-CHEST}~\cite{10368353} was recently proposed as a perturbation-based scheme that identifies relevant pilot and data subcarriers by observing the \ac{BER} sensitivity of the model to induced noise. Model-specific methods, by contrast, exploit the model's internal state, such as activations and gradients at each layer, to attribute the output to specific input features and internal components, rather than inferring importance from input-output behaviour alone.


\Ac{GRACE}~\cite{10621232} is a notable application of gradient-based interpretability to channel estimation, employing the basic \ac{LRP}~\cite{bach2015pixel} zero-rule ($LRP_z$) to explain \ac{FNN} estimators operating on single frequency-domain \ac{OFDM} symbols. While GRACE demonstrates that relevance-guided input filtering improves \ac{BER} performance, several limitations motivate the present work. First, prior analysis has focused on single-channel environments, providing no insight into how learned representations generalise across diverse propagation conditions. Second, analysing \ac{OFDM} symbols in isolation rather than entire frames makes relevance estimates sensitive to the specific channel realisation and noise sample of that symbol rather than reflecting stable underlying feature importance. Third, the basic rule does not account for modern architectural features such as residual connections or weight normalisation, restricting its applicability to simple FNN architectures. We also note that the $LRP_z$ rule is numerically unstable when pre-activations are close to zero, a condition that arises routinely in ReLU networks.

GRACE was developed to address input-level attribution, which leaves unaddressed a complementary and arguably more informative question: how the network internally represents the channel, and whether that internal representation is shared or fragmented across different propagation conditions. To the best of our knowledge, no prior work has examined the internal filter-level representations of a multi-channel-trained estimator with the aim of identifying structural redundancies and characterising the representational mechanism responsible for cross-channel generalisation. This work uses filter-level attribution both to characterise the network's internal representations and to guide architecture compression, producing a model whose relevance structure is transparent at both the input and filter levels and whose parameter and FLOP count are reduced for resource-constrained vehicular deployment.

\subsection{STRUCTURED PRUNING OF CONVOLUTIONAL NETWORKS}

Structured pruning removes entire filters or channels from a network, yielding hardware-efficient architectures compatible with standard dense matrix multiplication libraries without requiring specialised sparse convolution kernels~\cite{li2017pruning, he2018soft, he2024structured, cheng2024survey}. \ac{LRP}-based relevance criteria have been applied to prune image classifiers, where Yeom \emph{et al.}~\cite{yeom2021pruning} report better accuracy retention than magnitude-based and gradient-based pruning criteria at equivalent compression ratios, attributing this to the conservation property of \ac{LRP}, which provides stable importance estimates without per-layer normalisation. However, the focus of that line of work is purely compression alone. The internal representational structure revealed by filter-level relevance, which underpins the present analysis of cross-channel generalisation, remains unexamined.

In the \ac{TCN} compression literature, Yuan \emph{et al.}~\cite{yuan2024pruned} proposed a pruned tree-structured \ac{TCN} for industrial process prediction, using weight normalisation scaling factors as channel importance indicators within each block, with block-level importance derived from fully connected layer weights. Their criterion is magnitude-based and operates on output-side weights, whereas REACH applies gradient-based attribution directly to the convolutional features and analyses how the resulting filter relevance is distributed across multiple propagation conditions. The cross-channel representational question that motivates the present work therefore does not arise in their setting.

The need for compression methods tailored to wireless tasks, rather than generic computer-vision heuristics, is emphasised in a recent survey of model compression for wireless networks~\cite{khan2026compression}, which argues that physical-layer models should be assessed with wireless-native metrics such as \ac{NMSE} and \ac{BER} against \ac{SNR} rather than generic proxies such as top-1 accuracy or \ac{FLOPs} alone. The same survey notes that structured sparsity, which removes whole filters or channels and preserves regular tensor shapes, translates into predictable speedups on deployment hardware, whereas unstructured masks rarely do. For channel estimation specifically, it identifies redundancy in residual blocks and sensitivity to aggressive compression in the first and last layers, and observes that robustness is usually validated only on a finite set of simulated channels, so generalisation to unseen propagation conditions remains an open question. Our work addresses these points directly: we evaluate compact estimators with \ac{NMSE} and \ac{BER} against \ac{SNR} across six channel models, reduce capacity through a structured filter-width reduction rather than unstructured masking, and test generalisation explicitly on three held-out \ac{OOD} channels.

\section{SYSTEM MODEL AND PRELIMINARIES}
\label{sec:system_model}
This section establishes the signal model, the channel models, and the channel estimator that REACH analyses in Sections~\ref{sec:reach}--\ref{sec:pruning}. Section~\ref{subsec:ofdm} specifies the IEEE 802.11p OFDM signal model. Section~\ref{subsec:complex_to_real} describes the complex-to-real input representation used by the deep learning estimator. Section~\ref{subsec:channel_models} lists the six vehicular channel models and the in-distribution/out-of-distribution split. Section~\ref{sec:dl_estimators} summarises the DPA-RDCNN architecture introduced in~\cite{ngorima_access2026}, focusing on the elements that the interpretability analysis of
Section~\ref{sec:reach} references.

\subsection{OFDM SIGNAL MODEL}
\label{subsec:ofdm}
We consider an IEEE 802.11p \ac{OFDM} system with 64 subcarriers, of which 52 are active (48 data and 4 pilots at MATLAB indices $\{7,21,32,46\}$) with a \SI{10}{\mega\hertz} channel bandwidth and 50 \ac{OFDM} symbols per frame. Data symbols are drawn from a 16QAM constellation with Gray-coded mapping. Coded transmission uses a rate-$\tfrac{1}{2}$ convolutional code with constraint length 7 \ac{FEC}, the standard configuration for the IEEE 802.11p physical layer. The received signal at subcarrier $k$ and symbol $n$ is:
\begin{equation}
    Y[k,n] = H[k,n]\,X[k,n] + W[k,n],
    \label{eq:ofdm}
\end{equation}
where $H[k,n]$ is the complex channel frequency response, $X[k,n]$ is the transmitted symbol, and $W[k,n]$ is \ac{AWGN} with variance $\sigma_w^2$. The estimation objective is to infer $\hat{H}[k,n]$ from the four pilot subcarriers and from \ac{DPA}-processed data symbol estimates. Throughout this paper, bold uppercase symbols denote matrices and higher-order tensors, bold lowercase symbols denote vectors, and plain italic symbols denote scalars and indexed scalar entries.

\subsection{COMPLEX-TO-REAL REPRESENTATION}
\label{subsec:complex_to_real}
The \ac{DPA} stage produces an initial channel estimate $\hat{H}_{\text{DPA}}[k,n]$ for every subcarrier $k$ and \ac{OFDM} symbol $n$ in the frame. This complex estimate is converted to a real-valued input tensor for the \ac{DL} estimator by interleaving the real and imaginary parts along the symbol axis, producing 100 real-valued entries per subcarrier while the 52 active subcarriers are preserved. The resulting input tensor $\mathbf{X}_{\text{in}} \in \mathbb{R}^{52 \times 100}$ holds $5\,200$ real-valued features per frame, with rows indexing subcarriers and columns indexing the interleaved real and imaginary values, consistent with the convention used in~\cite{ngorima_access2026}. The ground-truth channel matrix $H[k,n]$ from~\eqref{eq:ofdm} is interleaved in the same way to form the regression target $\mathbf{H}_{\text{tgt}} \in \mathbb{R}^{52 \times 100}$, so that the network learns a mapping $\mathcal{N}: \mathbf{X}_{\text{in}} \mapsto \hat{\mathbf{H}} \in \mathbb{R}^{52 \times 100}$ on a common real-valued grid.

\subsection{VEHICULAR CHANNEL MODELS}
\label{subsec:channel_models}
Table~\ref{tab:channels} lists the six vehicular channel models used in this study, drawn from the Acosta-Marum empirical characterisation~\cite{acosta2007six}. The first three serve as the \ac{ID} training set and the remaining three are held out entirely from training for \ac{OOD} evaluation, supporting the assessment of cross-channel generalisation under realistic deployment conditions.

\begin{table}[ht!]
\centering
\caption{Vehicular channel models and training split.}
\label{tab:channels}
\begin{tabular}{lcccc}
\toprule
Channel  & Type & Doppler (Hz) & Delay (ns) & Split \\
\midrule
VTV-SDWW & ID   & 1200 & 302 & Train \\
RTV-SUS   & ID   &  600 & 701 & Train \\
VTV-EX   & ID   & 1200 & 401 & Train \\
VTV-UC   & OOD  &  500 & 401 & Evaluation only \\
RTV-UC   & OOD  &  300 & 501 & Evaluation only \\
RTV-EX   & OOD  &  700 & 401 & Evaluation only \\
\bottomrule
\end{tabular}
\end{table}

\subsection{DPA-RDCNN ARCHITECTURE}
\label{sec:dl_estimators}


The base estimator analysed in this work is the DPA-RDCNN architecture introduced in~\cite{ngorima_access2026}, where its design, hyperparameter optimisation, and benchmarking against classical and recurrent baselines are reported in full. To support the interpretability analysis that follows, we summarise here only the architectural elements that are explicitly referenced in Sections~\ref{sec:reach}--\ref{sec:pruning}.

The DPA-RDCNN processes complete \ac{OFDM} frames through $N_b = 5$ stacked residual blocks, each containing five weight-normalised one-dimensional dilated convolutional layers with kernel size $k = 4$, dilations $d \in \{1, 2, 4, 8, 16\}$, and hidden channel dimension $K = 256$. The first block
ingests the input tensor $\mathbf{X}_{\text{in}} \in \mathbb{R}^{52 \times 100}$ defined in Section~\ref{subsec:complex_to_real}, so that the residual block input $\mathbf{x}^{(b)}$ and output $\mathbf{y}^{(b)}$ are both in $\mathbb{R}^{52 \times K}$ for $b \geq 1$. For each residual block $b \in \{0, 1, \ldots, N_b - 1\}$, the output is

\begin{equation}
    \mathbf{y}^{(b)} = \mathrm{ReLU}\!\left(
        \underbrace{f^{(b)}\!\left(\mathbf{x}^{(b)}\right)}_{\text{convolutional path}}
        +
        \underbrace{g^{(b)}\!\left(\mathbf{x}^{(b)}\right)}_{\text{skip path}}
    \right),
    \label{eq:residual_block}
\end{equation}
where $f^{(b)}$ is the dilated convolutional stack and $g^{(b)}$ is a $1 \times 1$ convolution for block~$0$, where the input channel dimension differs from $K$, or the identity for blocks~$1$ to $4$. Fig.~\ref{fig:dpa_rdcnn_arch} shows the overall arrangement of the residual blocks and
the convolutional and skip paths inside each block. The baseline model has $6\,484\,580$ trainable parameters and performs $0.6729$~GFLOPs per inference frame.

Training follows the multi-channel mixed-\ac{SNR} strategy of~\cite{ngorima_access2026}, in which the network is trained jointly on the three \ac{ID} channels at nine \ac{SNR} levels spanning 0 to \SI{40}{\decibel} in \SI{5}{\decibel} steps. The cross-channel generalisation behaviour observed under this strategy is what REACH analyses at both the input and the filter level.

\section{REACH METHODOLOGY}
\label{sec:reach}

REACH applies gradient-based attribution to the multi-channel-trained DPA-RDCNN at two complementary levels: input feature attribution, which identifies the time-frequency components the model relies on, and internal filter attribution, which characterises the convolutional features that support cross-channel generalisation. Both levels share a common attribution rule based on the magnitude of the elementwise product of activations and their gradients, a formulation that handles residual connections, weight normalisation, and \ac{ReLU} non-linearities through standard automatic differentiation~\cite{ancona2017unified, simonyan2013deep}. In contrast to GRACE~\cite{10621232}, which applies the basic \ac{LRP} zero-rule to individual \ac{OFDM} symbols in a single channel model, REACH constructs per-environment and per-\ac{SNR} relevance maps by aggregating gradient-based attributions across $B = 500$ frames (Section~\ref{subsec:aggregation}), then combines these maps under the criteria of Section~\ref{subsec:feature_cat} at the input level and Section~\ref{subsec:filter_reach_method} at the filter level to produce stable estimates for multi-channel-trained models.

\subsection{ATTRIBUTION RULE}
\label{subsec:attr_rule}

REACH adopts the gradient$\times$input attribution rule~\cite{shrikumar2017learning, ancona2017unified}. Attribution is computed with respect to a scalar function $\phi(\mathbf{X}_{\text{in}}) = \|\hat{\mathbf{H}}\|_F^2$ of the network output, where $\|\cdot\|_F$ denotes the Frobenius norm. This function summarises the prediction without requiring access to the target $\mathbf{H}_{\text{tgt}}$ during the attribution stage. For a single frame with input $\mathbf{X}_{\text{in}} \in \mathbb{R}^{52 \times 100}$, the relevance per-frame at grid position $(k, i)$ is
\begin{equation}
    R(k, i) = \left| X_{\text{in}}(k, i) \right| \cdot 
              \left| \frac{\partial \phi(\mathbf{X}_{\text{in}})}{\partial X_{\text{in}}(k, i)} \right|,
    \label{eq:grad_x_input}
\end{equation}
which can be viewed as a first-order Taylor approximation to the change in $\phi$ produced by removing the input value at $(k, i)$. The use of absolute values discards sign information and produces a non-negative relevance score suitable for ranking and thresholding, consistent with prior work that uses gradient-based saliency for feature selection~\cite{simonyan2013deep}. Equation~\eqref{eq:grad_x_input} defines relevance for a single input frame; the per-environment and per-\ac{SNR} relevance maps used throughout REACH are obtained by aggregating this per-frame quantity over $B = 500$ independent frames, as specified in Section~\ref{subsec:aggregation}.

This formulation is preferred over the LRP-$\epsilon$ rule for two reasons. The $\epsilon$ stabiliser used by LRP-$\epsilon$ addresses the small-denominator instability of the basic $LRP_z$ rule in feedforward networks, but in residual architectures the issue persists in a different form. First, the residual addition in~\eqref{eq:residual_block} sums the convolutional and skip path outputs before the non-linearity, so the LRP-$\epsilon$ redistribution at this junction divides each path's relevance by $f^{(b)}(\mathbf{x}^{(b)}) + g^{(b)}(\mathbf{x}^{(b)}) + \epsilon$. This sum can vanish either through sign cancellation between the two paths, or through deep-fading conditions in which both paths carry low energy simultaneously, leaving the relevance scores dominated by the $\epsilon$ stabiliser rather than by the network's actual computation. Gradient$\times$input avoids this redistribution entirely, since the chain rule propagates gradients through the residual sum trivially and the parallel paths never share a denominator. Second, the weight-normalised convolutional layers in DPA-RDCNN introduce a reparametrisation not addressed by either $LRP_z$ or LRP-$\epsilon$, whereas gradient$\times$input handles such reparametrisations as part of the same chain-rule computation.

\subsection{TEMPORAL AND BATCH AGGREGATION}
\label{subsec:aggregation}

Single-frame attribution is dominated by the specific channel realisation and noise sample of that frame, producing high-variance relevance estimates. REACH suppresses this variance through multi-dimensional aggregation. For the input tensor $\mathbf{X}_{\text{in}} \in \mathbb{R}^{52 \times 100}$ defined in Section~\ref{subsec:complex_to_real}, the relevance map $\mathbf{R}_b \in \mathbb{R}^{52 \times 100}$ for frame $b$ has entries $R_b(k,i)$, where $k \in \{1, \ldots, 52\}$ is the subcarrier index and $i \in \{1, \ldots, 100\}$ is the index along the interleaved real and imaginary axis. REACH aggregates across $B = 500$ independent frames drawn from a specified channel model and \ac{SNR} level,
\begin{equation}
    \bar{R}(k, i) = \frac{1}{B} \sum_{b=1}^{B} R_b(k, i),
    \label{eq:batch_agg}
\end{equation}
yielding stable relevance estimates $\bar{\mathbf{R}} \in \mathbb{R}^{52 \times 100}$ for each channel model and \ac{SNR} combination.

\subsection{INPUT FEATURE CATEGORISATION}
\label{subsec:feature_cat}

Cross-channel important features are identified by combining the aggregated relevance maps from all six channel models. Let $\bar{R}^{(e)}(k, i)$ denote the aggregated relevance map for channel model $e \in \{1, \ldots, 6\}$, and let $P_{75}^{(e)}$ denote the 75th percentile of $\bar{R}^{(e)}$ taken over all $(k, i)$ pairs. The 75th percentile is a fixed analytical convention chosen for interpretability rather than a value optimised against estimation accuracy, and it provides a consistent quartile-based rule for separating features that are reliably high in relevance from the remainder, which is sufficient to characterise the cross-channel structure examined in this section. The accuracy and compression trade-off is instead established empirically in Section~\ref{sec:pruning}, where a range of filter widths is swept and the operating point is read off from the resulting NMSE curves rather than fixed in advance by a percentile cut. A feature $(k, i)$ is classified as \emph{cross-channel important} when it ranks in the top quartile of every channel model for at least one \ac{SNR} level:
\begin{equation}
    \bar{R}^{(e)}(k, i) \geq P_{75}^{(e)} 
    \quad \forall\, e \in \{1, \ldots, 6\}.
    \label{eq:cross_channel_def}
\end{equation}
Requiring top-quartile rank in all six channel models, rather than in a subset, ensures that the feature carries relevance information shared across the full set of training and held-out conditions, rather than information specific to one or two environments.
The set of SNR levels considered in this analysis is \{0, 20, 40\}\,\si{\decibel}, providing low, mid, and high-\ac{SNR} coverage of the training range. Computing relevance maps at all nine training SNRs would multiply the attribution cost without changing the qualitative cross-channel structure observed at the three sampled levels.

\subsection{FILTER-LEVEL ATTRIBUTION}
\label{subsec:filter_reach_method}

The input-level attribution rule of~\eqref{eq:grad_x_input} extends directly to intermediate convolutional features. Let $\mathbf{A}^{(b)} \in \mathbb{R}^{B \times K \times L}$ denote the output of the convolutional path $f^{(b)}$ of residual block $b$ in~\eqref{eq:residual_block}, where $B$ is the batch size, $K = 256$ is the hidden channel dimension, and $L = 52$ is the subcarrier dimension. The relevance of an individual element of this tensor is
\begin{equation}
    R^{(b)}_{n, f, l} = 
        \left| A^{(b)}_{n, f, l} \right| \cdot 
        \left| \frac{\partial \phi}{\partial A^{(b)}_{n, f, l}} \right|,
    \label{eq:elementwise_filter_relevance}
\end{equation}
which applies the attribution rule of~\eqref{eq:grad_x_input} to an intermediate tensor in place of the input tensor. Each element is captured by registering a forward hook to extract $A^{(b)}_{n, f, l}$ and a backward hook to extract the gradient through the same tensor in a single forward--backward pass.

A scalar relevance score for filter $f$ in block $b$ is obtained by averaging the elementwise relevance over the batch and subcarrier dimensions,
\begin{equation}
    R^{(b)}_f = 
        \frac{1}{B \cdot L} \sum_{n=1}^{B} \sum_{l=1}^{L} R^{(b)}_{n, f, l},
    \label{eq:filter_relevance}
\end{equation}
yielding a vector $\mathbf{r}^{(b)} \in \mathbb{R}^{K}$ that ranks the $K = 256$ filters in block $b$ by their contribution to the network output. Because $\mathbf{A}^{(b)}$ is captured \emph{before} the residual addition in~\eqref{eq:residual_block}, the score~\eqref{eq:filter_relevance} attributes relevance to the convolutional path independently of the skip path that bypasses it.



\subsubsection{FILTER TAXONOMY}
\label{subsubsec:taxonomy}

For each block $b$ and channel model $e$, the per-channel model relevance vector $\mathbf{R}^{(b, e)} \in \mathbb{R}^{K}$ is computed by restricting the average in~\eqref{eq:filter_relevance} to frames drawn from channel model $e$. Per-channel model thresholds $\tau^{(e)}_{75}$ and $\tau^{(e)}_{25}$ are then computed as the 75th and 25th percentiles of $\mathbf{R}^{(b, e)}$. Filters are classified into three categories:
\begin{align}
    \text{Universal:} \quad & R^{(b, e)}_f \geq \tau^{(e)}_{75} \;\; 
        \forall\, e, \label{eq:universal} \\
    \text{Redundant:} \quad & R^{(b, e)}_f \leq \tau^{(e)}_{25} \;\; 
        \forall\, e, \label{eq:redundant} \\
    \text{Environment-specific:} \quad & \text{otherwise.}
        \label{eq:specific}
\end{align}
Universal filters carry the cross-channel representation that supports operation on propagation conditions not seen during training. Environment-specific filters carry intermediate-relevance information that distinguishes channel conditions, the mechanistic interpretation of which is left for future work. Redundant filters can be removed without expected performance loss, providing the principled compression budget exploited in Section~\ref{sec:pruning}.

\section{ATTRIBUTION RESULTS}
\label{sec:results}

This section reports the attribution results at both the input and filter levels. Section~\ref{subsec:setup} describes the experimental setup that all subsequent results build on. Section~\ref{sec:input_reach} presents input-level relevance analysis, and Section~\ref{sec:filter_reach} presents filter-level relevance analysis.

\subsection{EXPERIMENTAL SETUP}\label{subsec:setup}

\subsubsection{DATASET GENERATION}
Datasets are generated in MATLAB R2022b using a baseband simulation of the IEEE~802.11p \ac{PHY}, adapted from the open-source framework of Gizzini \emph{et al.}~\cite{gizzini2020deep}. Each of the six vehicular channel models in Table~\ref{tab:channels} is instantiated using MATLAB's \texttt{comm.RayleighChannel} object with time-varying tap coefficients driven by Jakes' Doppler spectrum, and each frame is transmitted through an independent channel realisation generated with a deterministic seed equal to the frame index. For each channel model, 18\,000 frames are generated for training and 2\,000 independently for testing, with the testing set evaluated at each $E_b/N_0$ point in $\{0, 5, \ldots, 40\}$\,dB. The multi-channel training set is the concatenation of the three \ac{ID} training sets, yielding 54\,000 frames spanning nine \ac{SNR} levels.

\subsubsection{BASELINE MODEL AND TRAINING PROTOCOL} 
The DPA-RDCNN baseline analysed throughout this work is the trained checkpoint released with~\cite{ngorima_access2026}, where hyperparameter selection and training are reported in full. The optimiser is Adam ($\beta_1{=}0.9$, $\beta_2{=}0.999$) with learning rate $\eta = 4.77 \times 10^{-4}$, batch size 128, and StepLR scheduling with step size 11 and decay factor 0.948. Dropout with $p = 0.228$ is applied after convolutional layers, weight decay $\sim 10^{-6}$ provides $L_2$ regularisation, and early stopping halts training after 20 consecutive epochs without validation loss improvement.

\subsubsection{COMPACT MODELS}
The compact DPA-RDCNN variants evaluated in Section~\ref{sec:pruning} are trained from scratch at each target filter width $K' \in \{210, 192, 160, 128, 96\}$, using the same optimiser, schedule, and stopping criterion as the baseline. Each configuration is trained with five independent random seeds to quantify training variance.

\subsubsection{EVALUATION PROTOCOL}
\ac{NMSE} and \ac{BER} are reported per channel model and per \ac{SNR} point, computed over the 2\,000 independent test frames of the corresponding channel. For attribution, relevance maps are aggregated over $B = 500$ frames per channel-and-\ac{SNR} combination (Section~\ref{subsec:aggregation}), drawn from the test partition. All models are implemented in PyTorch and trained on NVIDIA A40 and A30 \acp{GPU}.

\subsection{INPUT-LEVEL}
\label{sec:input_reach}
This section examines which input features the network relies on across the six channel models. Two questions follow: whether the relevance patterns are shared across channel models, and whether the identified features are sufficient on their own to drive accurate estimation.

\subsubsection{CROSS-CHANNEL RELEVANCE PATTERNS}

\begin{figure}[ht!]
\centering
\includegraphics[width=0.75\linewidth]%
    {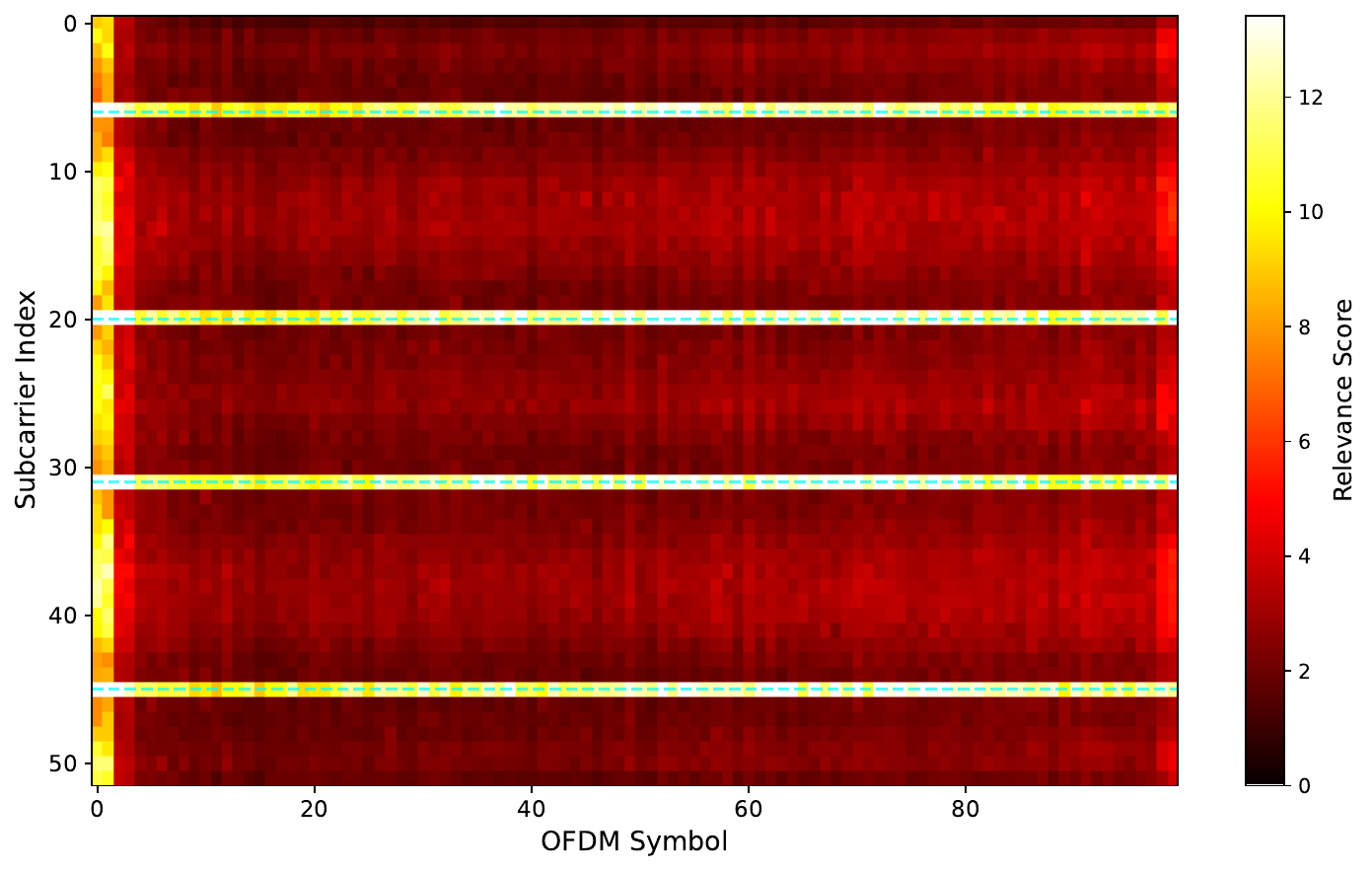}
\caption{Input relevance map for DPA-RDCNN on \ac{VTV-SDWW} at 0\,\si{\decibel} \ac{SNR}, computed from~\eqref{eq:batch_agg}. The horizontal axis indexes the 100 interleaved real and imaginary values across the 50 OFDM symbols of the frame (two columns per symbol), and the vertical axis indexes the 52 subcarriers. The colour bar gives the relevance score in arbitrary units.}
\label{fig:tcn_example_heatmap}
\end{figure}

Fig.~\ref{fig:tcn_example_heatmap} shows an example relevance map for the \ac{VTV-SDWW} channel model at 0\,\si{\decibel} \ac{SNR}, obtained by applying the aggregation rule in~\eqref{eq:batch_agg}.

\begin{figure}[ht!]
\centering
\includegraphics[width=0.70\linewidth]%
    {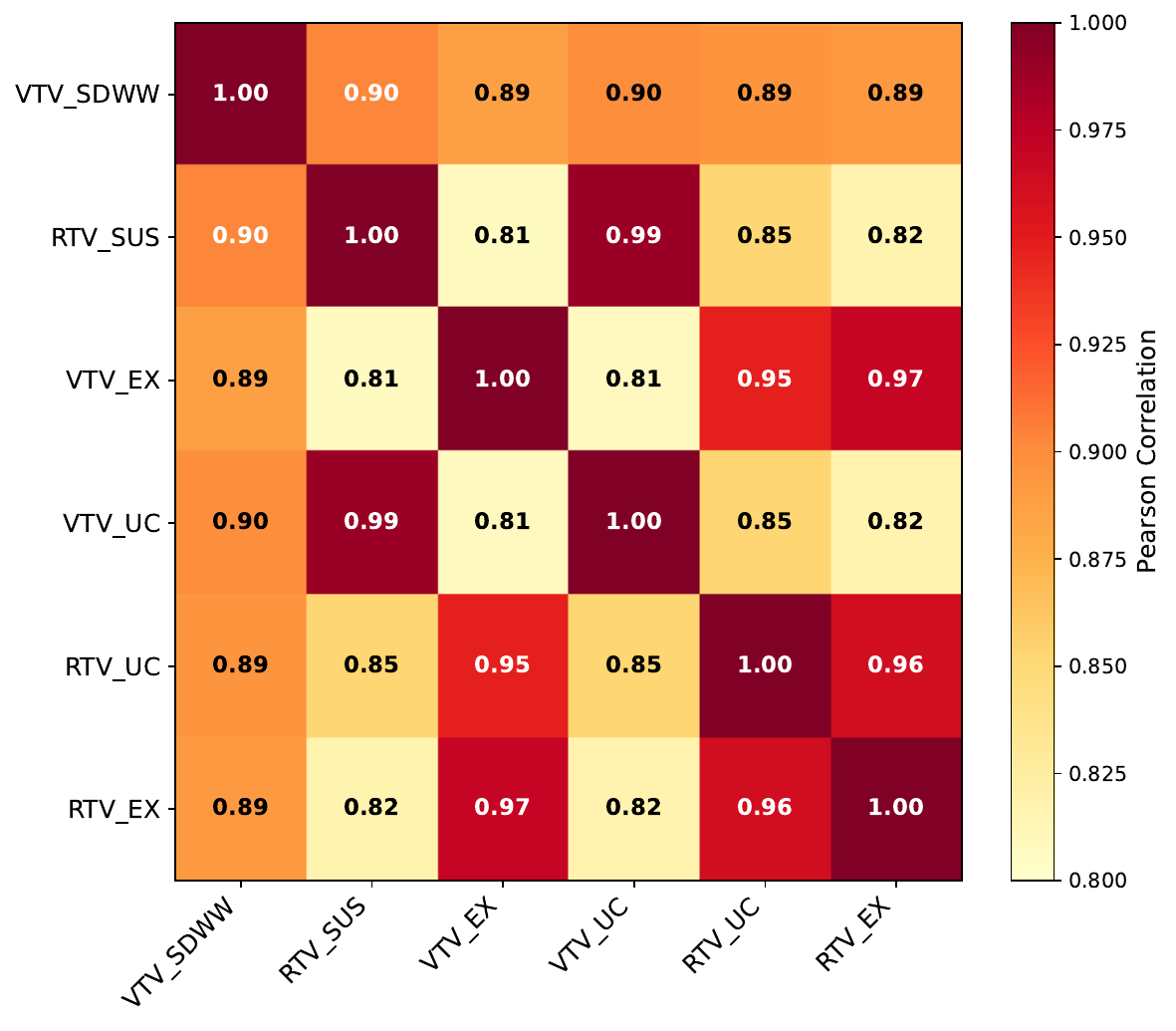}
\caption{Input-level cross-channel relevance correlation matrix at 20\,\si{\decibel} \ac{SNR}. All pairwise correlations exceed 0.81, including between \ac{ID} and \ac{OOD} channel models, indicating that the network relies on similar time--frequency patterns across all propagation environments.}
\label{fig:tcn_correlation_matrix}
\end{figure}

Relevance exhibits distinct patterns. The four pilot subcarriers form bright horizontal bands across the full frame. The first interleaved columns carry elevated relevance across all subcarriers: as the earliest data symbols in the DPA recursion, they are seeded most directly by the preamble-based initial estimate. Between the pilot bands, relevance rises with distance from the nearest pilot, lowest immediately adjacent and highest midway between. Elsewhere, it is approximately uniform and low. This pattern recurs across all six channel models and all \ac{SNR} levels, varying in magnitude rather than location, motivating the cross-channel similarity quantified in Fig.~\ref{fig:tcn_correlation_matrix}.

All pairwise correlations exceed 0.81, including those between the three \ac{ID} channel models and the three \ac{OOD} channel models held out from training. This provides an initial interpretability-based account of the observed \ac{OOD} generalisation: the network does not rely on input patterns specific to any one channel model, but on a common time-frequency structure shared across all evaluated propagation environments.

\subsubsection{CROSS-CHANNEL IMPORTANT FEATURES}

\begin{figure}[ht!]
\centering
\includegraphics[width=0.75\linewidth]%
    {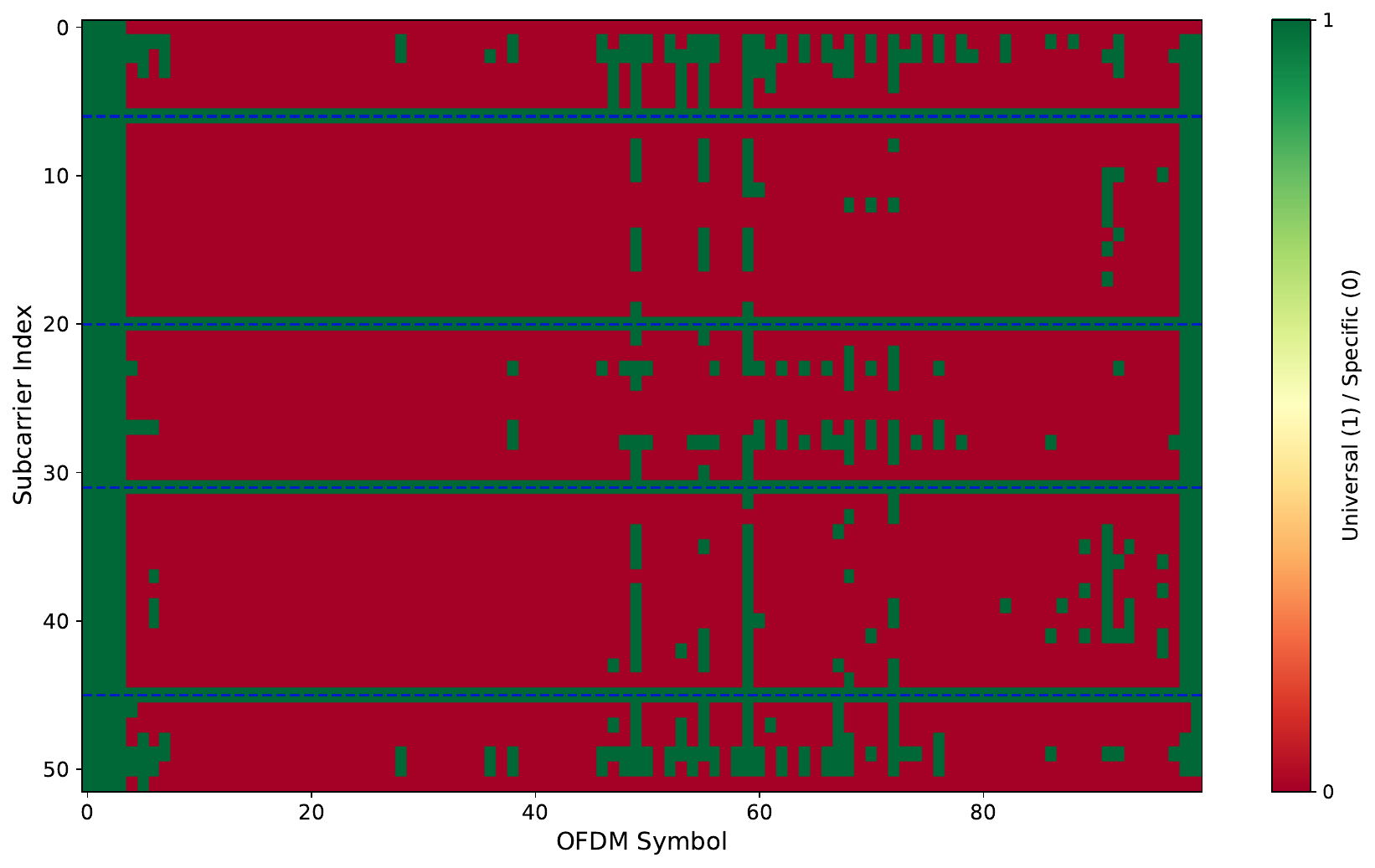}
\caption{Cross-channel important features in the $52 \times 100$ input space, identified by the criterion in~\eqref{eq:cross_channel_def} over SNR levels $\{0, 20, 40\}$\,\si{\decibel}. The horizontal axis indexes the 100 interleaved real and imaginary values across the 50 OFDM symbols (two columns per symbol), and the vertical axis indexes the 52 subcarriers. Green indicates features classified as cross-channel important, and red indicates features that do not meet the criterion.}
\label{fig:tcn_universal_features}
\end{figure}

Applying the criterion in~\eqref{eq:cross_channel_def} across SNR levels $\{0, 20, 40\}$\,\si{\decibel}, 1\,020 of the 5\,200 input features (19.6\%) are cross-channel important. Fig.~\ref{fig:tcn_universal_features} shows the resulting binary mask, in which each cell is the real or imaginary component of one OFDM symbol at one subcarrier, so that each OFDM symbol occupies two adjacent columns.

The mask is binary rather than continuous because the criterion in~\eqref{eq:cross_channel_def} tests whether the relevance exceeds the 75th-percentile threshold in every propagation environment, not the relevance magnitude itself. The four pilot subcarriers at indices $\{6, 20, 31, 45\}$ (equivalently the MATLAB indices {7, 21, 32, 46} of Section~\ref{subsec:ofdm}) form fully green horizontal bands, reflecting their role as channel reference anchors important in every channel model. The first few columns are also fully green, marking the earliest OFDM symbols as important across all subcarriers, and a further cluster appears mid-frame, spanning a wide range of subcarriers rather than the pilots alone.

Table~\ref{tab:universal_statistics} quantifies this distribution by symbol region, a span of OFDM symbols along the horizontal axis, and by spectral region, a span of subcarriers along the vertical axis. All four pilot subcarriers are cross-channel important (400/400). The remaining 620 features fall on data subcarriers, consistent with the role of \ac{DPA} refinement: the network relies on the pilots in every condition and additionally recruits data positions where the \ac{DPA}-refined estimates carry the most informative residual channel information. The early-frame region carries a disproportionate share of this data contribution, with 42.8\% of the symbols 0--5 region cross-channel important against 8.1\% of the symbols 6--15 region, reflecting the proximity of the earliest OFDM symbols to the preamble-based initial estimate.


%
\begin{table}[t!]
\centering
\caption{Distribution of cross-channel important features (union over SNR levels $\{0, 20, 40\}$\,\si{\decibel}). Counts are over interleaved features (the real and imaginary component of each symbol at each subcarrier); symbol regions span the OFDM-symbol axis and spectral regions span the subcarrier axis.}
\label{tab:universal_statistics}
\begin{tabular}{lcc}
\toprule
\textbf{Feature Region} & \textbf{Total}
  & \textbf{Cross-Channel (\%)} \\
\midrule
\multicolumn{3}{l}{\textit{Symbol regions}} \\
Symbols 0--5 (early)       & 624  & 267 (42.8\%) \\
Symbols 6--15 (mid-early)  & 1\,040 &  84 (8.1\%)  \\
Symbols 16--30 (mid)       & 1\,560 & 283 (18.1\%) \\
Symbols 31--49 (late)      & 1\,976 & 386 (19.5\%) \\
\midrule
\multicolumn{3}{l}{\textit{Spectral regions}} \\
Edge SC (0--5, 47--51)     & 1\,100 & 226 (20.5\%) \\
Mid SC (6--20, 32--46)     & 3\,000 & 569 (19.0\%) \\
Central SC (21--31)        & 1\,100 & 225 (20.5\%) \\
\midrule
\multicolumn{3}{l}{\textit{Subcarrier type}} \\
Pilot subcarriers          &  400 & 400 (100\%)  \\
Data subcarriers           & 4\,800 & 620 (12.9\%) \\
\midrule
\textbf{Total}             & 5\,200 & 1\,020 (19.6\%) \\
\bottomrule
\end{tabular}
\end{table}
%

\subsubsection{PILOT VS DATA SUBCARRIER CONTRIBUTIONS}

\begin{figure*}[ht!]
\centering
\includegraphics[width=\linewidth]%
    {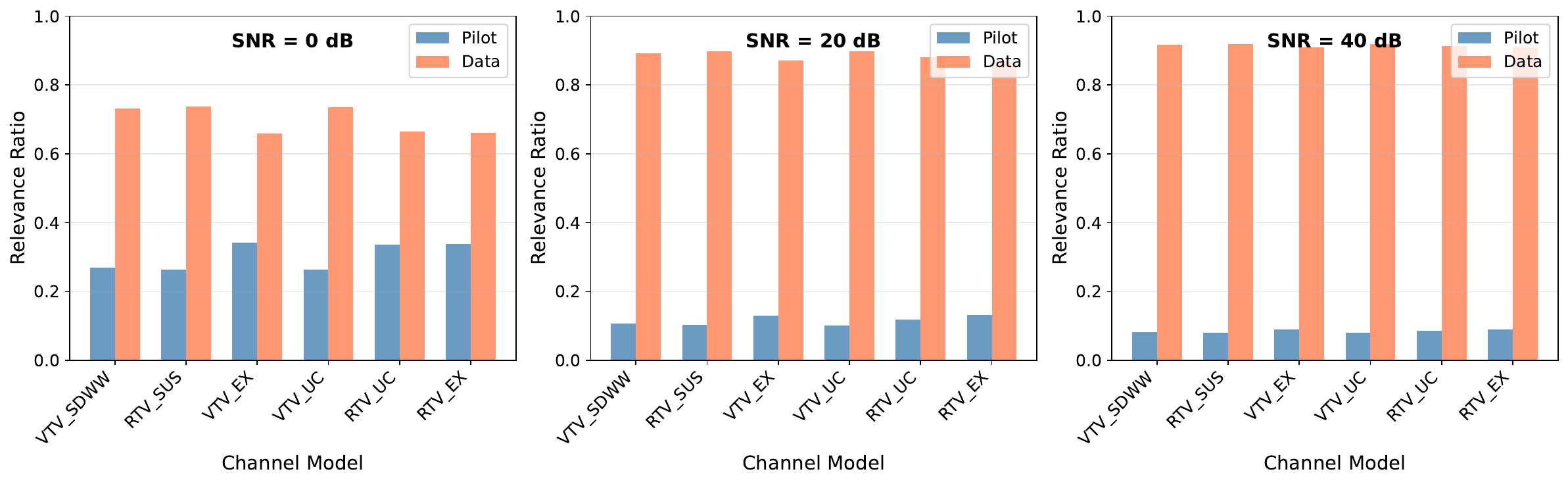}
\caption{Relative relevance contributions of pilot and data subcarriers across channel models and \ac{SNR} levels. Data subcarriers contribute 65--92\% of total relevance, and their contribution grows with \ac{SNR} as \ac{DPA}-refined estimates become more reliable. The pattern is consistent across all six channel models.}
\label{fig:tcn_pilot_data}
\end{figure*}

Fig.~\ref{fig:tcn_pilot_data} shows the relevance carried by the pilot and data subcarriers at each channel model and \ac{SNR} level, each expressed as a fraction of the total relevance at that operating point, so that the pilot and data shares sum to one.
At low \ac{SNR}, pilots provide the dominant reference anchors for estimation. As \ac{SNR} increases and decision-directed data estimates become more reliable, the model shifts reliance toward the denser spectral coverage of data positions, effectively increasing the virtual pilot density. This \ac{SNR}-adaptive behaviour is consistent across all six channel models, including the three \ac{OOD} channels.

\subsubsection{CROSS-CHANNEL FEATURE VALIDATION}

\begin{figure*}[ht!]
\centering
\includegraphics[width=0.85\linewidth]%
    {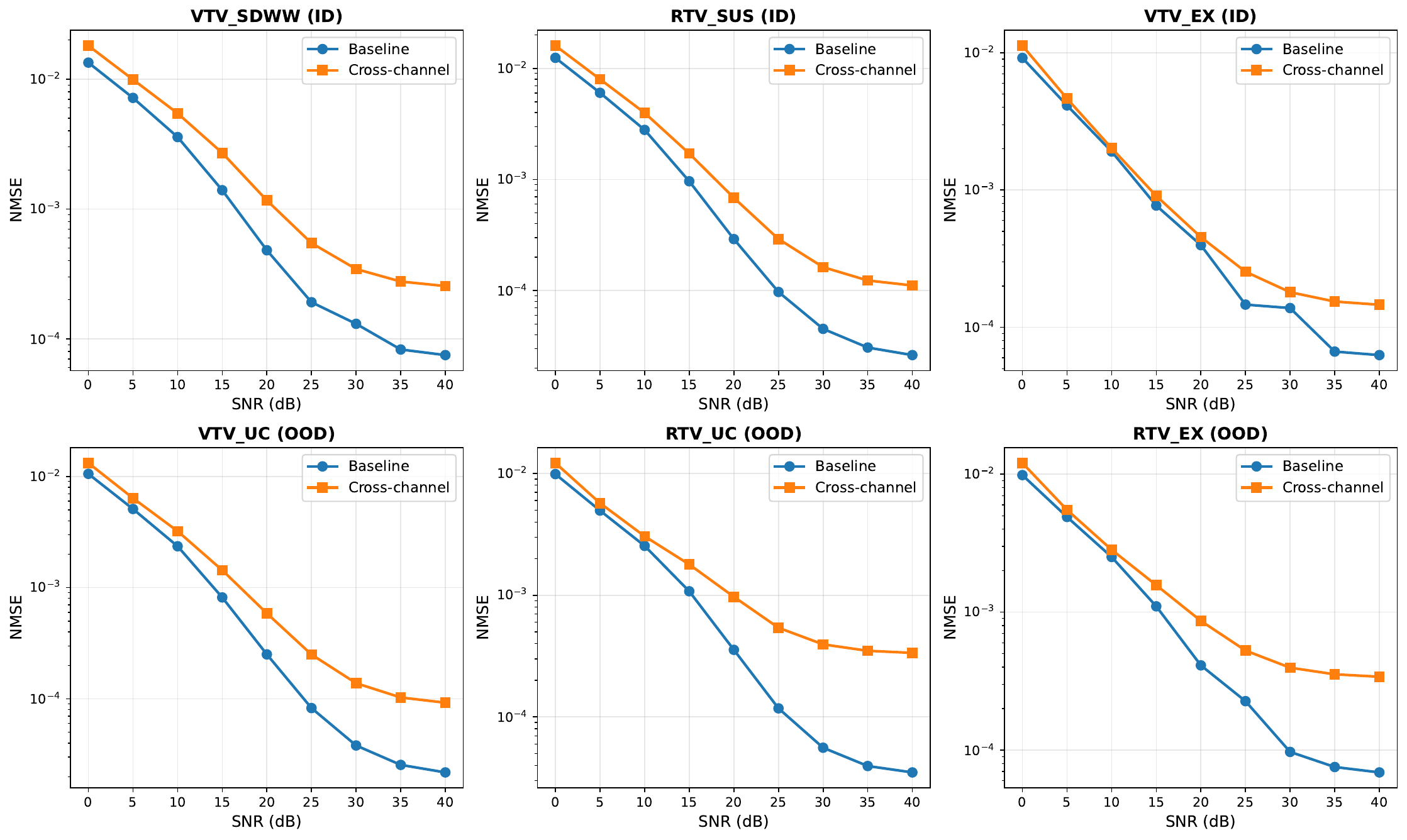}
\caption{\ac{NMSE} comparison between the baseline (5\,200 features) and the cross-channel model (1\,020 features, 19.6\%) on \ac{ID} and \ac{OOD} channel models.}
\label{fig:tcn_id_performance}
\end{figure*}

To verify that the 1\,020 cross-channel important features are functionally sufficient for accurate estimation, we train a \emph{cross-channel model} using only these features. Training uses the same three \ac{ID} channels, nine \ac{SNR} levels (0--40\,\si{\decibel} in 5\,\si{\decibel} steps), and training hyperparameters as the baseline of Section~\ref{subsec:setup}, with a single random seed. The real and imaginary components of each complex value are masked independently, each retained or zeroed according to its own entry in the cross-channel mask, so the real part of a position may be kept while the imaginary part is zeroed. Retained positions keep their original values during training and inference, and all other positions are set to zero throughout. The resulting cross-channel model is then compared against the baseline. Fig.~\ref{fig:tcn_id_performance} shows \ac{NMSE} against \ac{SNR} for all six channel models.

\begin{figure*}[ht!]
    \centering
    \includegraphics[width=0.85\textwidth]{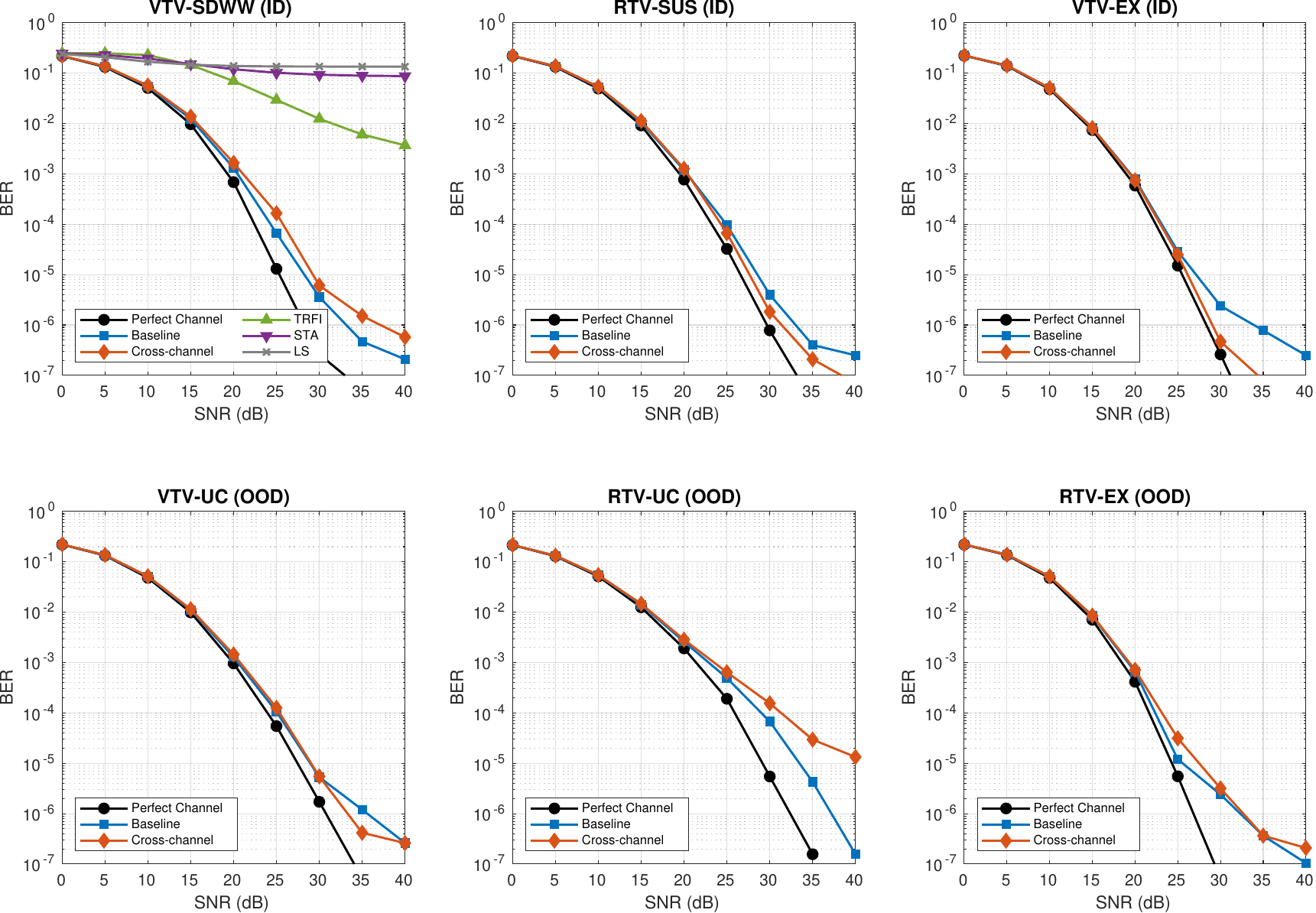}
    \caption{\ac{BER} performance: baseline against the cross-channel model (19.6\% of features) across all six channel models. On the VTV-SDWW panel the classical LS, STA and TRFI estimators are included for comparison with established practical methods.  The cross-channel model closely tracks the baseline up to 25\,\si{\decibel}, indicating functional sufficiency of the identified feature set for practical deployments.}
\label{fig:ber_universal_vs_Baseline}
\end{figure*}

Fig.~\ref{fig:tcn_id_performance} compares the \ac{NMSE} of the baseline against the cross-channel model on all six channel models. Across the practical \ac{SNR} range (0--25\,\si{\decibel}), the cross-channel model remains within 2--4\,\si{\decibel} of the baseline on all six channel models. Above 25 dB the gap widens as the baseline approaches its estimation floor more closely, with the largest deviations observed on the \ac{OOD} channels at 40\,\si{\decibel}. Fig.~\ref{fig:ber_universal_vs_Baseline} compares \ac{BER} performance across all six channel models. Up to 25\,\si{\decibel}, the cross-channel model closely tracks the baseline. In several cases (\ac{VTV-EX}, \ac{RTV-SUS}) the cross-channel model slightly outperforms the baseline at higher \ac{SNR}. A regularisation effect from removing low-relevance components is documented in the filter pruning literature~\cite{he2018soft}, and provides one candidate explanation, although the conditions under which the effect arises here are not investigated in detail in this work.

On the VTV-SDWW channel we additionally include three classical estimators, least squares (LS), spectral temporal averaging (STA), and time-domain reliable frequency-domain interpolation (TRFI), as a representative comparison against estimators that remain in practical use [3]. Across the full SNR range the strongest classical estimator, TRFI, does not fall below a BER of approximately $4 x 10^-3$, while LS and STA remain close to $10^-1$. Both the baseline and the cross-channel model fall several orders of magnitude lower, confirming that the deep learning estimator substantially outperforms the classical baselines on this channel.

\subsection{FILTER-LEVEL}
\label{sec:filter_reach}

The input-level analysis of Section~\ref{sec:input_reach} established that the network relies on a common time--frequency input pattern across all six channel models. This section turns to the internal representation. Two questions follow: whether different channel models activate a shared internal filter population, and whether the resulting taxonomy supports a principled compression strategy. The analysis proceeds in four steps: a block-level relevance measurement that tests whether any residual block can be removed wholesale, a filter taxonomy that classifies every convolutional filter as universal, environment-specific, or redundant, a dilation-level breakdown that probes the temporal granularity of the attribution, and a cross-channel correlation that quantifies how similar the filter recruitment is across the six channel models.

\subsubsection{BLOCK-LEVEL RELEVANCE}
\label{subsec:block_relevance}

\begin{figure*}[!t]
\centering
\resizebox{\linewidth}{!}{%
\begin{tikzpicture}[
    >=stealth,
    node distance=6mm and 8mm,
    box/.style={draw, rounded corners=1pt, minimum height=9mm, 
                minimum width=16mm, align=center, font=\footnotesize},
    block/.style={draw, rounded corners=2pt, minimum height=12mm, 
                  minimum width=22mm, align=center, fill=blue!8, 
                  font=\footnotesize\bfseries},
    op/.style={draw, circle, inner sep=0pt, minimum size=5mm, font=\footnotesize},
    arrow/.style={->, thick},
    skip/.style={->, thick, dashed, gray}
]

\node[box] (input) {Input\\$\mathbf{X}_{\text{in}}$\\$52 \times 100$};

\node[block, right=of input] (b1) {Residual\\Block 1};
\node[block, right=of b1]    (b2) {Residual\\Block 2};
\node[block, right=of b2]    (b3) {Residual\\Block 3};
\node[block, right=of b3]    (b4) {Residual\\Block 4};
\node[block, right=of b4]    (b5) {Residual\\Block 5};

\node[op, right=of b5]       (sum) {$+$};
\node[box, right=of sum]     (head) {$1\times1$\\Conv\\Head};
\node[box, right=of head]    (out)  {Output\\$\hat{\mathbf{H}}$\\$52 \times 100$};

\draw[arrow] (input) -- (b1);
\draw[arrow] (b1) -- (b2);
\draw[arrow] (b2) -- (b3);
\draw[arrow] (b3) -- (b4);
\draw[arrow] (b4) -- (b5);
\draw[arrow] (b5) -- (sum);
\draw[arrow] (sum) -- (head);
\draw[arrow] (head) -- (out);

\coordinate (skipA) at ([yshift=12mm]input.north);
\coordinate (skipB) at ([yshift=12mm]sum.north);
\node[font=\scriptsize\itshape, above=0mm] at ($(skipA)!0.5!(skipB)$)
    {global skip ($1\times1$ conv)};
\draw[skip] (input.north) -- (skipA) -- (skipB) -- (sum.north);

\node[font=\footnotesize\itshape, below=10mm of b3] (zlabel) 
    {Inside each residual block:};

\node[box, below=10mm of zlabel] (zin) {$\mathbf{x}^{(b)}$};
\node[block, right=of zin, minimum width=35mm, minimum height=18mm] (zconv)
    {Conv path $f^{(b)}$:\\5 dilated 1D convs\\$d \in \{1,2,4,8,16\}$};
\node[op, right=of zconv] (zsum) {$+$};
\node[box, right=of zsum] (zout) {ReLU\\$\mathbf{y}^{(b)}$};

\draw[arrow] (zin) -- (zconv);
\draw[arrow] (zconv) -- (zsum);
\draw[arrow] (zsum) -- (zout);

\coordinate (zskipA) at ([yshift=10mm]zin.north);
\coordinate (zskipB) at ([yshift=10mm]zsum.north);
\node[font=\scriptsize\itshape, above=0mm] at ($(zskipA)!0.5!(zskipB)$)
    {skip path $g^{(b)}$};
\draw[skip] (zin.north) -- (zskipA) -- (zskipB) -- (zsum.north);

\end{tikzpicture}%
}
\caption{Schematic of the DPA-RDCNN architecture. The five residual blocks are connected sequentially with a global skip route (top) that bypasses the entire block stack. The inset (bottom) shows the internal structure of a single residual block: a convolutional path $f^{(b)}$ containing 
five dilated 1D convolutions, and a parallel skip path $g^{(b)}$. The convolutional path is the subject of the filter-level analysis in Section~\ref{sec:filter_reach}.}
\label{fig:dpa_rdcnn_arch}
\end{figure*}
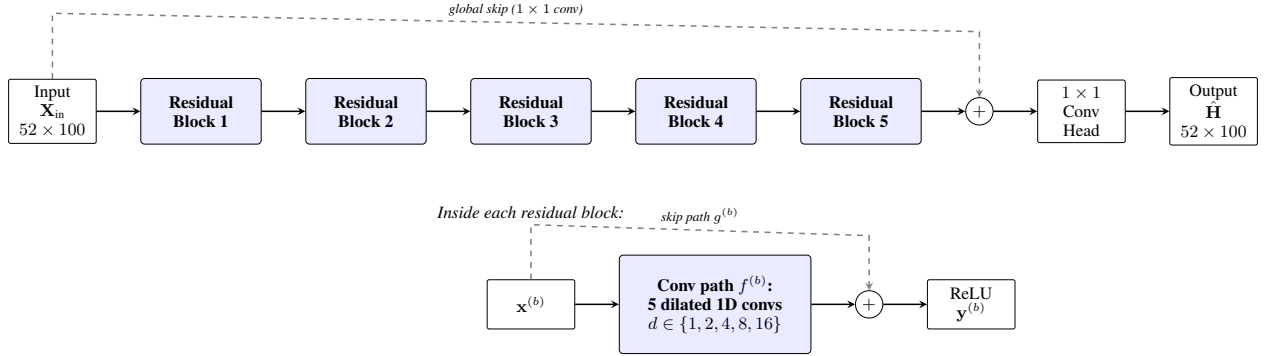

The first question concerns the relative contribution of each residual block. In the DPA-RDCNN schematic of Fig.~\ref{fig:dpa_rdcnn_arch}, each block routes information through a convolutional path $f^{(b)}$ and a parallel skip path $g^{(b)}$ that contribute jointly to the block's output. To isolate the contribution of the learned convolutions, we compute the per-block relevance contribution by averaging the filter-level relevance scores of~\eqref{eq:filter_relevance} across all filters in the block and normalising across blocks. The result indicates how much each block's convolutional path contributes to the prediction relative to the others, independently of the skip route that bypasses it.

\begin{figure}[ht!]
    \centering
    \includegraphics[width=0.70\columnwidth]{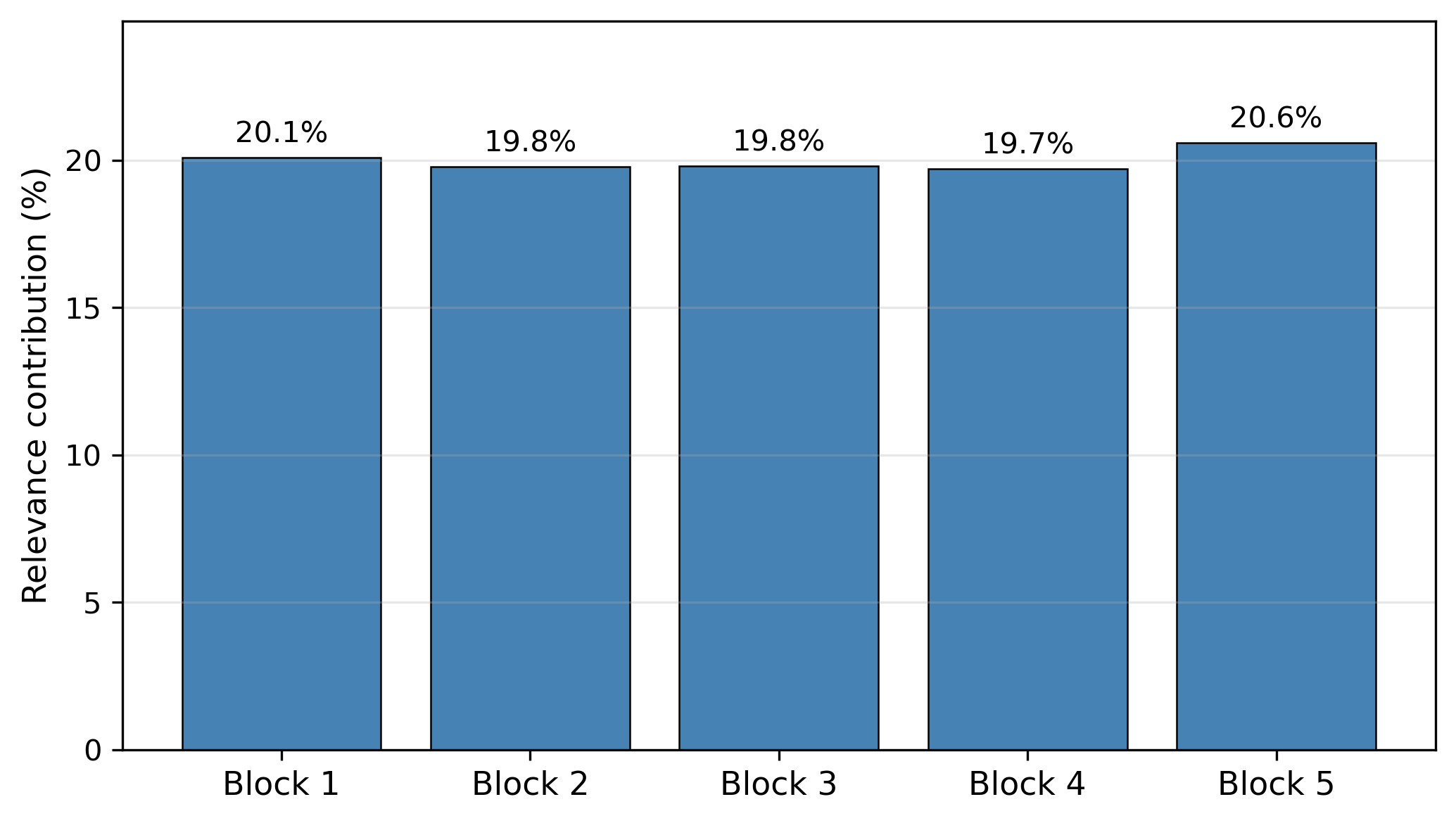}
    \caption{Convolutional-path relevance contribution per residual block of the DPA-RDCNN, averaged across all six channel models and three \ac{SNR} levels.}
    \label{fig:block_relevance}
\end{figure}

Fig.~\ref{fig:block_relevance} shows the convolutional-path relevance contribution per block, averaged across all six channel models and three \ac{SNR} levels.

All five blocks contribute between 19.7\% and 20.6\% of the total, within one percentage point of uniform distribution. This contrasts with the layer-wise importance variation commonly reported for image classification networks, where shallower or deeper layers are pruned more aggressively~\cite{he2024structured, cheng2024survey}. Two consequences follow. First, block-level removal is not a viable compression strategy, since each block contributes approximately one fifth of the total prediction capacity. Second, the network operates in a distributed mode in which every block participates approximately equally, motivating the filter-level analysis that follows.

\subsubsection{FILTER TAXONOMY}
\label{subsec:taxonomy}
 
\begin{figure}[ht!]
    \centering
    \includegraphics[width=0.70\columnwidth]{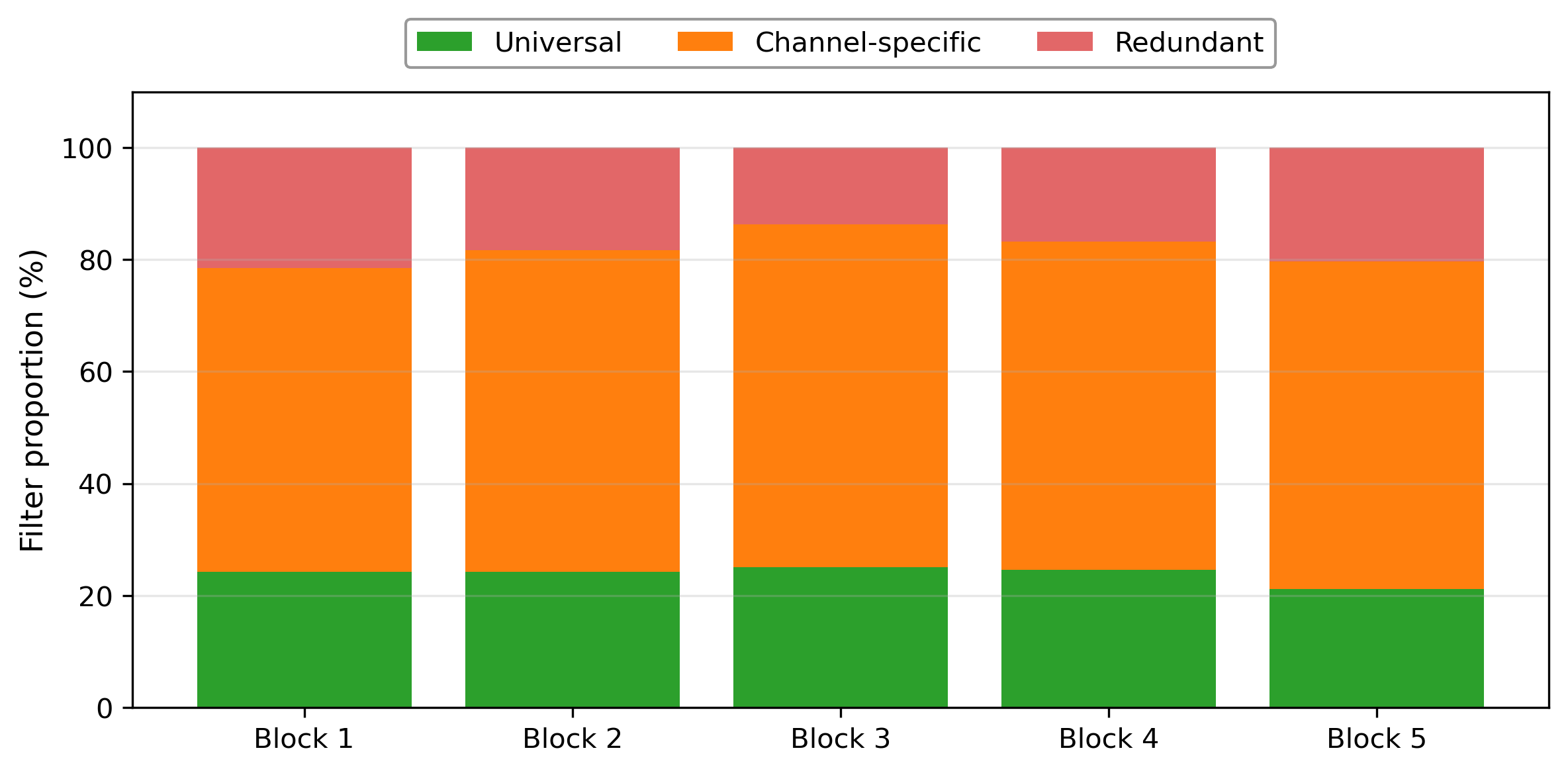}
    \caption{Filter taxonomy per residual block. The composition is consistent across all five blocks at approximately 23.8\% universal (green), 58.0\% environment-specific (orange), and 18.1\% redundant (red).}
    \label{fig:filter_taxonomy}
\end{figure}

Applying the classification criteria in~\eqref{eq:universal}--\eqref{eq:specific} to the per-block, per-channel-model filter relevance scores $R^{(b, e)}_f$ computed from 500 frames per channel-and-\ac{SNR} combination yields the taxonomy summarised in Table~\ref{tab:taxonomy} and visualised in Fig.~\ref{fig:filter_taxonomy}.

\begin{table}[ht!]
\centering
\caption{Filter taxonomy across all five blocks ($5 \times 256 = 1\,280$ filters total).}
\label{tab:taxonomy}
\begin{tabular}{lccc}
\toprule
Category               & Count & Fraction & Prunable \\
\midrule
Universal              &  305  & 23.8\%   & No  \\
Environment-specific   &  743  & 58.0\%   & No  \\
Redundant              &  232  & 18.1\%   & Yes \\
\midrule
Total                  & 1\,280  & 100\%    &     \\
\bottomrule
\end{tabular}
\end{table}


Three observations follow from Table~\ref{tab:taxonomy} and Fig.~\ref{fig:filter_taxonomy}. First, the composition of each block is nearly identical, supporting the distributed processing structure identified in Section~\ref{subsec:block_relevance}: the network has not specialised individual blocks for different aspects of the estimation problem. Second, the 58\% of filters classified as environment-specific have intermediate relevance in every condition, falling neither uniformly high nor uniformly low. Whether this population encodes Doppler-adaptive or delay-spread-adaptive information is left for future work using probing or ablation methods. Third, the 18.1\% of filters classified as redundant have consistently low relevance across all six channel models and three \ac{SNR} levels, corresponding to 232 filters that can be removed without expected performance loss. This constitutes the principled compression budget exploited in Section~\ref{sec:pruning}.

\subsubsection{DILATION CONTRIBUTION}
\label{subsec:dilation}

\begin{figure*}[ht!]
    \centering
    \includegraphics[width=\linewidth]{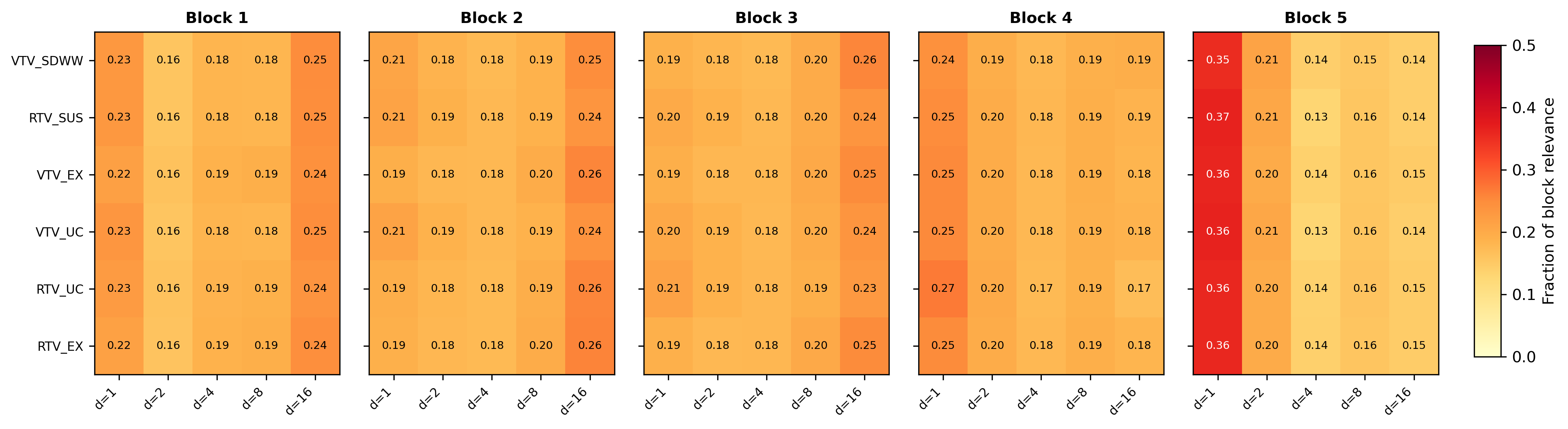}
    \caption{Normalised dilation contribution per block and channel model.}
    \label{fig:dilation}
\end{figure*}

Fig.~\ref{fig:dilation} shows the contribution of each dilation level within each block, expressed as a fraction of the block's total relevance and annotated with the value in each cell. Within a block of five dilations, an even split corresponds to 0.20 per dilation.

Blocks 1 to 3 are close to this even split, with every dilation between 0.16 and 0.26 and no single dilation dominating. From block 4 onward, the distribution tilts toward the shortest dilation: block 4 places 0.25 of its relevance at $d = 1$, and block 5 concentrates 0.36 at $d = 1$ against a mean of 0.16 across the four longer dilations. This $d = 1$ concentration in block 5 is consistent across all six channel models (0.35 to 0.37), including the three \ac{OOD} channels. The network therefore distributes relevance evenly across temporal scales in the early blocks and shifts toward short-range processing in the final block, where the $d = 1$ convolution carries roughly twice the relevance of an even split.

\subsubsection{CROSS-CHANNEL FILTER CORRELATION}
\label{subsec:correlation}
\begin{figure*}[ht!]
    \centering
    \includegraphics[width=\linewidth]{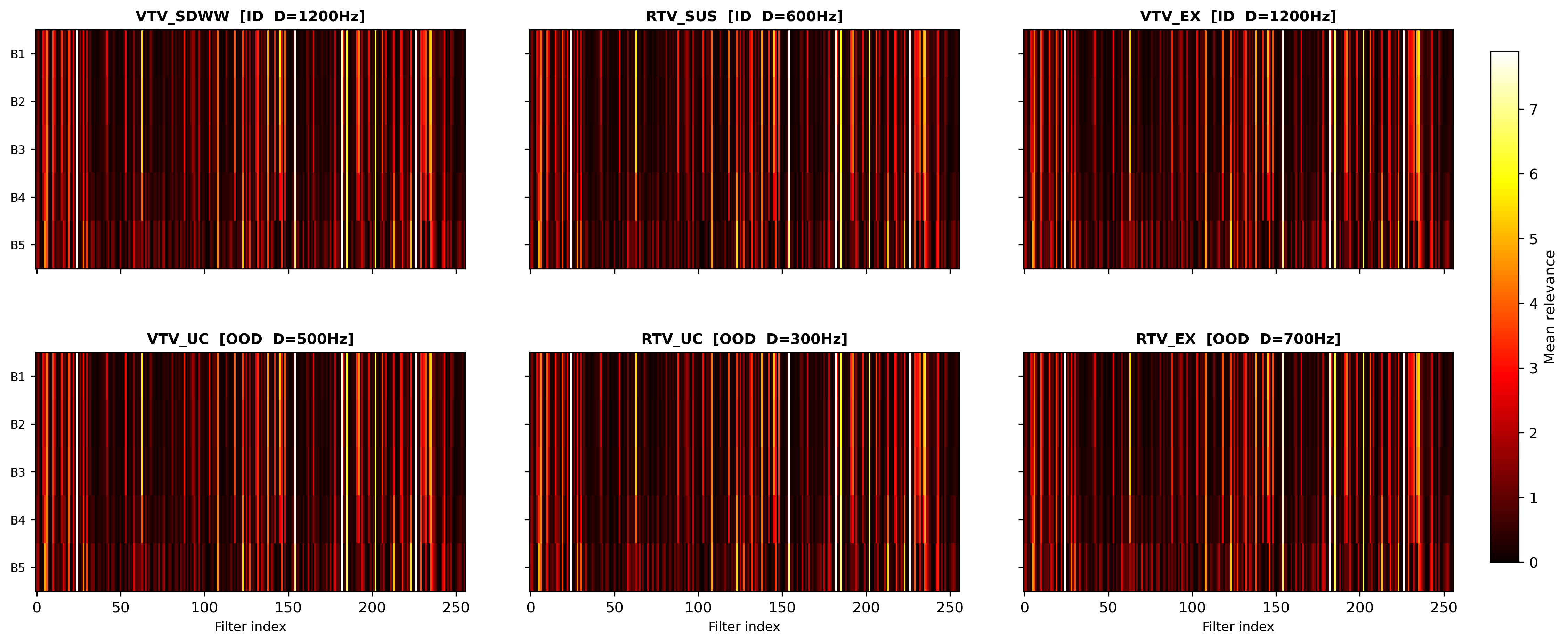}
    \caption{Filter relevance maps for the five residual blocks and all six channel models, including the three \ac{OOD} channels held out from training. High-relevance filters concentrate at similar indices across the panels, with on average 95.6\% of the 50 highest-relevance filters per block common to all six channel models.}
    \label{fig:filter_per_channel}
\end{figure*}

The taxonomy of Section~\ref{subsec:taxonomy} indicates that all six channel models recruit a similar population of filters. Fig.~\ref{fig:filter_per_channel} shows the per-channel-model filter relevance maps, in which the high-relevance filters appear at similar indices across the six channel models, including the three \ac{OOD} conditions held out from training. This consistency is difficult to verify by eye at this scale, so we quantify it directly: of the 50 highest-relevance filters in each block, 95.6\% on average are common to all six channel models, and 96.8\% of the 125 highest-relevance filters across the whole network are shared by every channel. The high-relevance filter indices are therefore largely the same across all propagation conditions, including the held-out ones.

To quantify the overall similarity, we flatten the filter relevance map of each channel model into a vector $\mathbf{r}^{(e)} \in \mathbb{R}^{N_b K}$, where $N_b = 5$ is the number of blocks and $K = 256$ is the number of filters per block, and compute the correlation between every pair of channel models. The mean pairwise Pearson correlation is $r = 0.9963$ (minimum 0.9933), and the mean Spearman rank correlation is 0.96 (minimum 0.93), confirming that both the magnitude and the ranking of filter relevance are near-identical across channel models. This is substantially higher than the input-level correlation reported in Section~\ref{sec:input_reach} (mean $r \approx 0.89$, minimum $r > 0.81$), revealing a structural asymmetry: input representations are moderately similar across channel models, but internal filter representations are near-identical. Multi-channel training does not produce separate internal subnetworks for each channel condition; instead it converges to a single shared filter representation, and the three \ac{OOD} channels activate this representation with relevance patterns very close to those of the three \ac{ID} channels on which the network was trained.

\begin{figure}[!t]
    \centering
    \includegraphics[width=0.75\columnwidth]{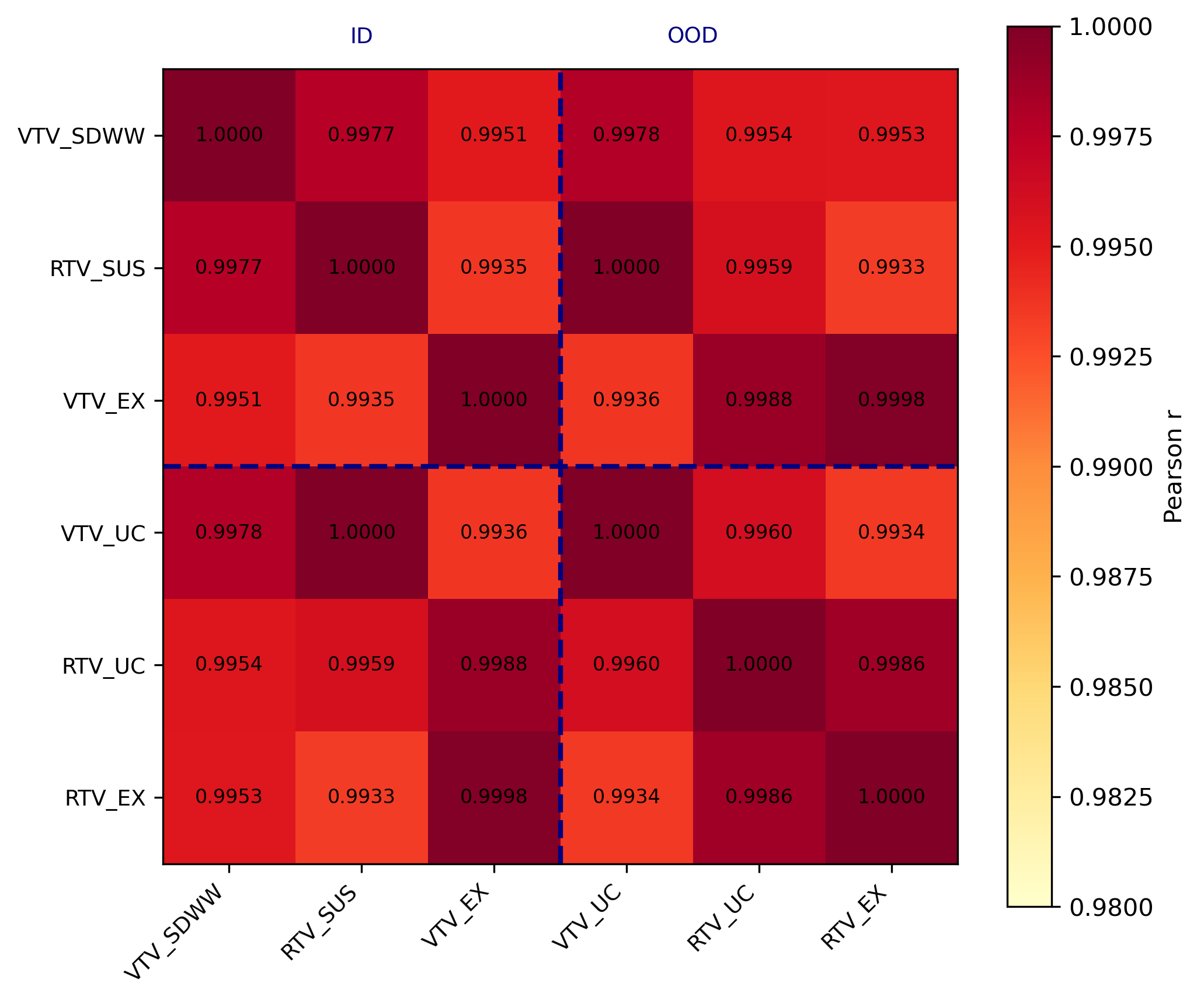}
    \caption{Cross-channel filter relevance correlation matrix. The colour map is restricted to the range 0.98--1.00 to resolve variation within the high-value region. All pairwise correlations exceed 0.99 with a mean $r = 0.9963$. The dashed lines separate the three \ac{ID} channel models (top-left) from the three \ac{OOD} channel models (bottom-right).}
    \label{fig:correlation}
\end{figure}

This consistency also supports a single-mask strategy for the architecture compression of Section~\ref{sec:pruning}: because the relevance ranking is stable across all six channel models, a single ranking computed by averaging across them covers every channel condition simultaneously. A fuller discussion of the mechanism implied by this asymmetry is deferred to Section~\ref{sec:discussion}.

\section{RELEVANCE-GUIDED ARCHITECTURE COMPRESSION}
\label{sec:pruning}

The filter taxonomy of Section~\ref{subsec:taxonomy} indicates that approximately 18\% of the filters in each residual block carry low relevance across all six channel models. This suggests that a network with a reduced number of filters per block should match the estimation accuracy of the full $K = 256$ baseline, provided the reduction is consistent with the relevance ranking. This section evaluates the suggestion empirically.

\subsection{STRATEGY AND ARCHITECTURAL CONSIDERATIONS}

We construct compact DPA-RDCNN variants by reducing the hidden filter count of every residual block from $K = 256$ to a smaller value $K'$, selected so as to retain at least the universal and environment-specific filters of the taxonomy. The hidden filter count, referred to as the \emph{filter width} hereafter, is the only architectural parameter varied across configurations. The number of residual blocks ($N_b = 5$), the number of dilated convolutions per block ($N_d = 5$), the kernel size ($k = 4$), the dilation pattern, and the input--output dimensions are held fixed.

The taxonomy of Section~\ref{subsec:taxonomy} shows that the network can be narrowed: the redundant filters can be removed without losing the universal and environment-specific filters that carry the estimation-relevant representation, which sets a target filter width $K'$ for each compact model. Surgical removal of individual filters from the trained baseline is not straightforward in this architecture, however. The residual addition in \eqref{eq:residual_block} requires the convolutional path output and the skip path output to share the same filter dimension. Removing filters from the convolutional path alone produces a shape mismatch, and the remaining weights are co-adapted to the components that have been removed, which typically degrades performance unless the network is retrained. We therefore train a self-consistent compact model at each target filter width $K'$ from scratch, using the same multi-channel mixed-\ac{SNR} data and training hyperparameters as the baseline. The taxonomy sets $K'$ rather than providing a surgical mapping from baseline filters to compact-model filters, and each configuration is trained with five independent random seeds to quantify training variance.

\subsection{FLOP ANALYSIS}

The FLOP count of a compact DPA-RDCNN with filter width $K'$ is
\begin{equation}
    \mathrm{FLOPs}(K') = 2 N_{\mathrm{SC}} \!\left[
        \underbrace{N_{\mathrm{in}} K'}_{\text{global skip}}
        + \sum_{b=0}^{N_b-1} \mathcal{F}_b(K')
        + \underbrace{K' N_{\mathrm{out}}}_{\text{output head}}
    \right],
    \label{eq:flops}
\end{equation}
with $\mathcal{F}_b(K') = N_{\mathrm{in}}^{(b)} K' k + (N_d - 1) K'^2 k + \mathbf{1}[b = 0] N_{\mathrm{in}} K'$, where $N_{\mathrm{in}}^{(b)} = N_{\mathrm{in}}$ for $b = 0$ and $K'$ otherwise, $N_{\mathrm{SC}} = 52$, $N_{\mathrm{in}} = N_{\mathrm{out}} = 100$, $k = 4$, and $N_d = 5$. The three terms in $\mathcal{F}_b$ are the first dilated convolution, the remaining $N_d - 1$ dilated convolutions, and the block-0 downsample that matches the input channel count to $K'$. The quadratic dependence on $K'$ in the per-block term dominates as $K'$ grows, so even a modest reduction in filter width produces a substantial reduction in the number of operations. At $K' = 256$, the uncompressed baseline, the count is 0.673~GFLOPs per frame, and at $K' = 160$ it is 0.267~GFLOPs.

\subsection{NMSE UNDER COMPRESSION}
\label{sec:compression_tradeoff}

Table~\ref{tab:compression} reports the parameter count, the FLOP count, and the NMSE degradation at 20\,\si{\decibel} \ac{SNR} for five compact configurations relative to the $K' = 256$ baseline.

\begin{table}[ht!]
\centering
\caption{Compression versus NMSE trade-off. $\Delta$NMSE values are averaged across all six channel models at 20\,\si{\decibel} \ac{SNR} and reported as mean $\pm$ standard deviation across five training seeds.}
\label{tab:compression}
\setlength{\tabcolsep}{5pt}
\begin{tabular}{crrrc}
\toprule
$K'$ & Parameters & 
       \makecell{Parameter\\reduction (\%)} & 
       \makecell{FLOPs\\(G)} & 
       \makecell{$\Delta$NMSE\\(\si{\decibel})} \\
\midrule
256 & 6\,484\,580 & ---  & 0.673 & ---                    \\
210 & 4\,392\,040 & 32.3 & 0.456 & $+0.16 \pm 0.45$       \\
192 & 3\,683\,812 & 43.2 & 0.382 & $+0.41 \pm 0.34$       \\
160 & 2\,578\,340 & 60.2 & 0.267 & $+0.58 \pm 0.36$       \\
128 & 1\,669\,476 & 74.3 & 0.173 & $+1.28 \pm 0.30$       \\
 96 &   957\,220  & 85.2 & 0.099 & $+2.06 \pm 0.18$       \\
\bottomrule
\end{tabular}
\end{table}

Two configurations are of particular interest. At $K' = 210$, which corresponds to retaining roughly the 81\% of filters identified as either universal or environment-specific by the taxonomy, the average NMSE degradation is $0.16 \pm 0.45$\,\si{\decibel}. The standard deviation across seeds exceeds the mean, so the degradation is statistically consistent with zero at this operating point. At $K' = 160$, which reduces the parameter count by 60.2\% and the FLOP count by 60.4\%, the average NMSE degradation is $0.58 \pm 0.36$\,\si{\decibel}. The degradation accelerates below $K' = 128$ (where it reaches $1.28 \pm 0.30$\,\si{\decibel}), which suggests $K' = 160$ as a practical lower bound when sub-1\,\si{\decibel} 
NMSE loss is required.

\subsection{BER UNDER COMPRESSION}
\label{sec:ber_pruning}

\begin{figure*}[ht!]
    \centering
    \includegraphics[width=0.85\textwidth]{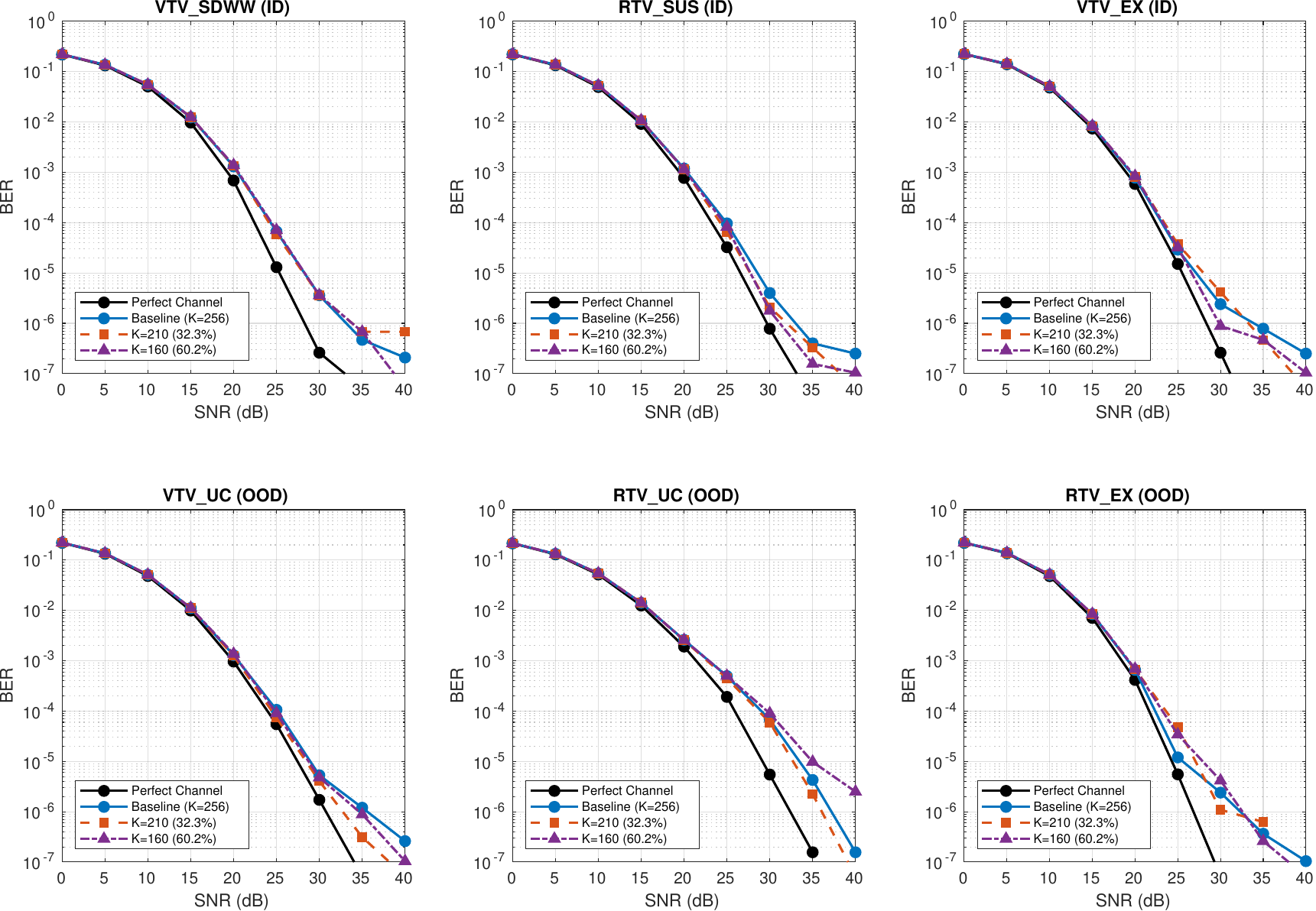}
    \caption{Coded BER of the baseline ($K' = 256$) and the two practical compact configurations ($K' = 210$, 32.3\% parameter reduction, and $K' = 160$, 60.2\% parameter reduction) across all six channel models. Top row: \ac{ID} channels. Bottom row: \ac{OOD} channels. The \emph{perfect channel} curve provides the BER lower bound under ideal \ac{CSI}.}
    \label{fig:ber_pruned}
\end{figure*}

Fig.~\ref{fig:ber_pruned} shows the coded BER of the baseline and the two practical operating points ($K' = 210$ and $K' = 160$) across all six channel models.

At $K' = 210$, the BER curves of the compact model lie essentially on top of the baseline across all six channel models, consistent with the near-zero NMSE degradation reported in Table~\ref{tab:compression}. At $K' = 160$, a small separation appears at high \ac{SNR}, but the BER curves remain close to the baseline throughout the practical \ac{SNR} range. The \ac{OOD} channels show a similar pattern to the \ac{ID} channels at both compression levels. The curve variability visible below a BER of roughly $10^{-6}$ on some panels reflects the finite test set rather than a difference between configurations: each BER point is estimated from 2\,000 test frames, about $1.9 \times 10^{7}$ coded bits, so a BER of $10^{-6}$ corresponds to only of the order of ten bit errors and the estimate becomes statistically unstable. The configurations should therefore be compared above this level, where the error counts are large enough for the BER estimates to be reliable.

\subsection{OOD GENERALISATION UNDER COMPRESSION}
\label{sec:ood_pruning}

\begin{figure*}[ht!]
    \centering
    \includegraphics[width=0.85\linewidth]{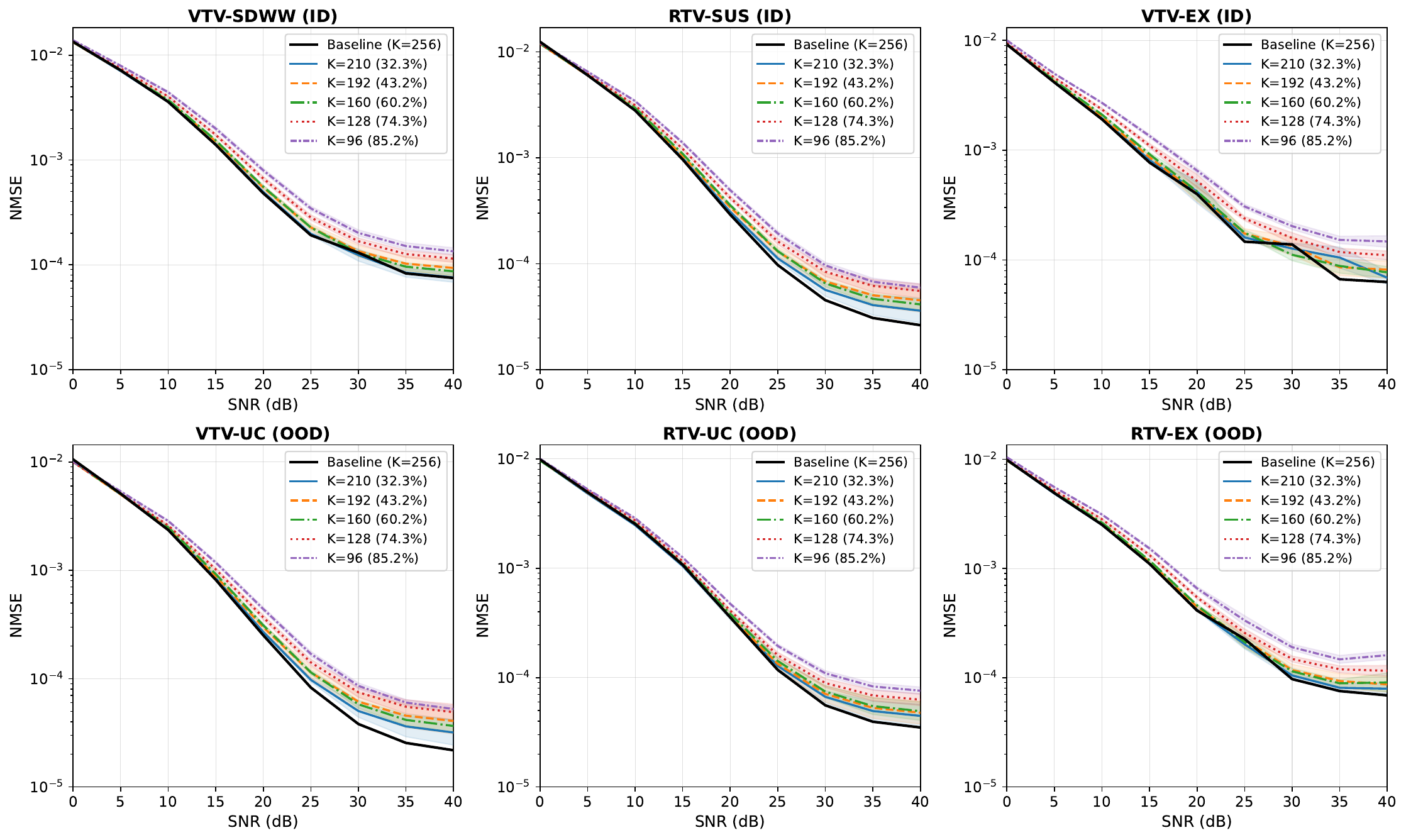}
    \caption{Per-channel NMSE against \ac{SNR} for every compact configuration, plotted as mean $\pm$ standard deviation across five seeds. Top row: \ac{ID} channels. Bottom row: \ac{OOD} channels.}
    \label{fig:nmse_all}
\end{figure*}

Table~\ref{tab:compression} reports the mean \ac{NMSE} degradation averaged over all six channel models. Splitting this average into the \ac{ID} and \ac{OOD} groups reveals a small but consistent pattern: at every compression level, the \ac{OOD} degradation is approximately equal to or slightly smaller than the \ac{ID} degradation. At $K' = 160$, the \ac{OOD} mean degradation is $0.57 \pm 0.15$\,\si{\decibel} compared with $0.59 \pm 0.48$\,\si{\decibel} on \ac{ID} channels. At $K' = 96$, the \ac{OOD} mean is $1.90 \pm 0.15$\,\si{\decibel} compared with $2.21 \pm 0.20$\,\si{\decibel} on \ac{ID} channels. The \ac{OOD} standard deviation is also smaller than the \ac{ID} standard deviation at every operating point, indicating that the seed-to-seed variability of the compact models is lower on the held-out channel models than on the training channel models. The reason for this asymmetry is not clear to us, and we leave its investigation for future work. Fig.~\ref{fig:nmse_all} shows the per-channel \ac{NMSE} curves with seed-to-seed confidence bands.

\subsection{JOINT INPUT AND FILTER COMPRESSION}
\label{sec:unified}

The preceding compression results vary the filter width $K'$ while the input remains the full 5\,200-feature representation. The input-level analysis of Section~\ref{sec:input_reach} suggested that 1\,020 of the 5\,200 features are sufficient to retain the bulk of the estimation accuracy. We now combine both reductions in a single configuration: a compact model with filter width $K' \in \{210, 192, 160, 128, 96\}$ trained on the 1\,020-feature masked input, where the remaining 4\,180 input positions are set to zero throughout training and inference.
This joint configuration is compared against two references: the full baseline (filter width $K' = 256$, full 5\,200-feature input) and the cross-channel model of Section~\ref{sec:input_reach} (filter width $K' = 256$, 1\,020-feature input).

\subsubsection{NMSE PERFORMANCE}

\begin{figure*}[ht!]
    \centering
    \includegraphics[width=0.85\linewidth]{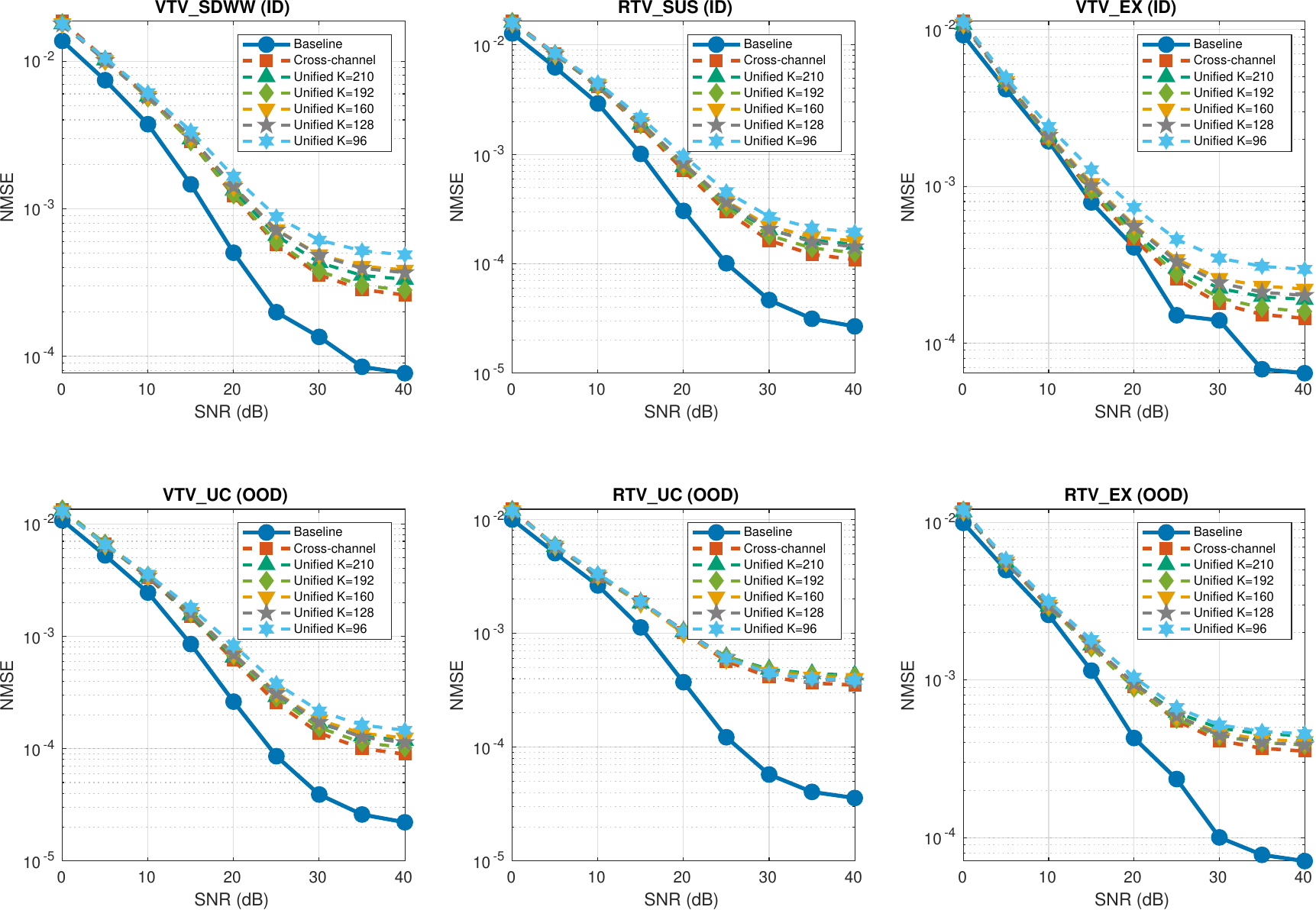}
    \caption{Per-channel NMSE for the joint input-masking and filter-width reduction configurations, compared with the full baseline and the input-only cross-channel model.}
    \label{fig:unified_nmse}
\end{figure*}

Fig.~\ref{fig:unified_nmse} shows the per-channel NMSE for the joint configurations and the two references.
Across the low-to-mid \ac{SNR} range, the joint configurations follow the cross-channel model closely, and the separation between the $K' = 210$ and $K' = 96$ joint configurations is small. This suggests that, once the input has been restricted to the 1\,020 features identified in Section~\ref{sec:input_reach}, the filter width is no longer the dominant constraint on \ac{NMSE}: the input mask sets a floor that the filter-width reduction does not breach. At high \ac{SNR}, the joint configurations approach the same estimation floor as the cross-channel model.


\subsubsection{BER}

\begin{figure*}[ht!]
    \centering
    \includegraphics[width=0.85\textwidth]{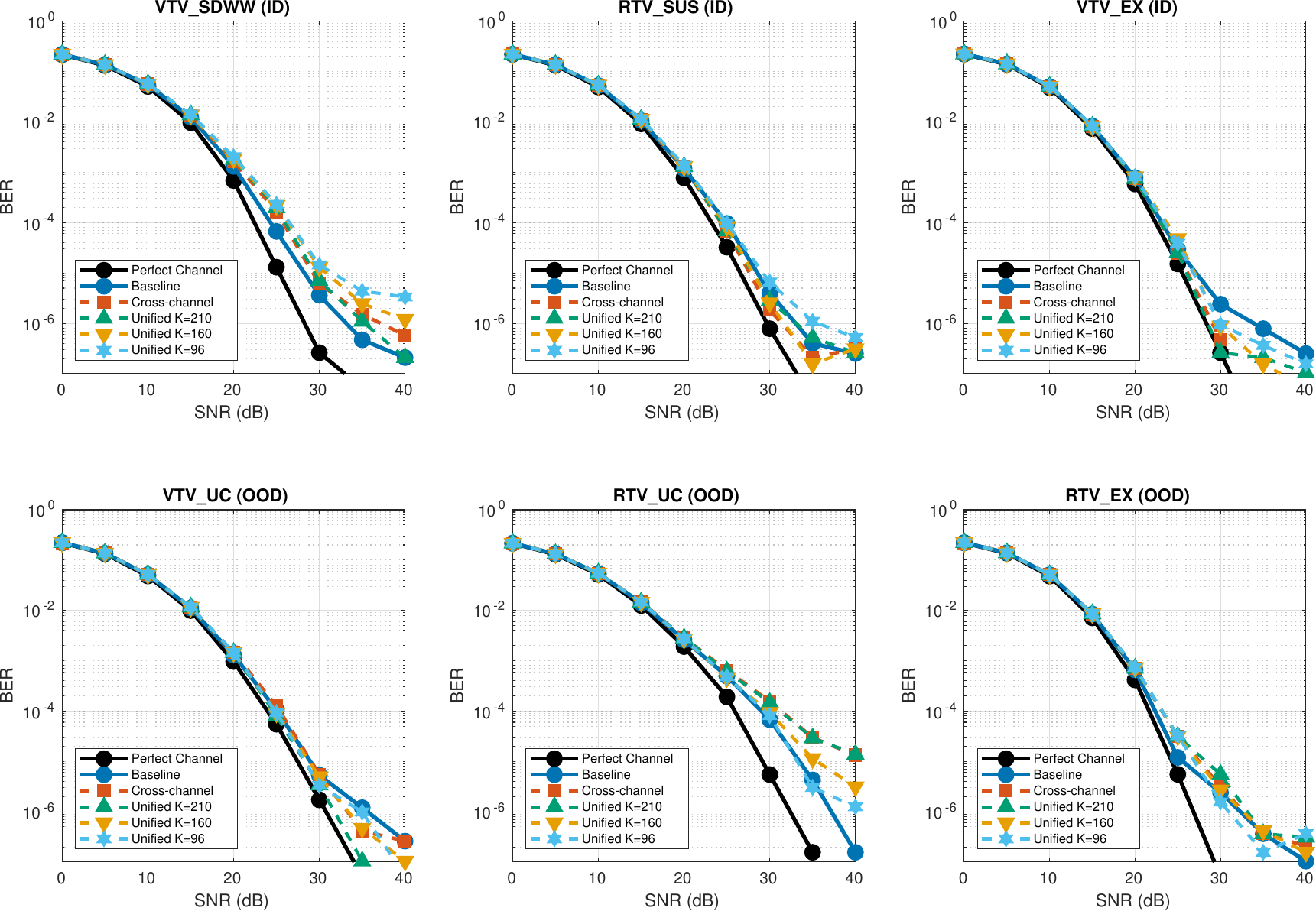}
    \caption{Coded \ac{BER} for the joint compression configurations ($K' = 210$ and $K' = 160$, 1\,020-feature input), the full baseline ($K' = 256$, 5\,200-feature input), and the cross-channel model ($K' = 256$, 1\,020-feature input). Top row: \ac{ID} channels. Bottom row: \ac{OOD} channels.}
    \label{fig:unified_ber}
\end{figure*}

Fig.~\ref{fig:unified_ber} shows the corresponding coded BER curves.
The joint $K' = 210$ and $K' = 160$ configurations track the cross-channel model across the practical \ac{SNR} range (0--25\,\si{\decibel}), with visible separation only above 30\,\si{\decibel} where the \ac{BER} falls below $10^{-5}$. Even at $K' = 96$, the \ac{BER} stays within one decade of the cross-channel model throughout the BER decay region. The \ac{OOD} channels follow the same pattern as the \ac{ID} channels.

\subsubsection{COMPRESSION ACHIEVED}

The parameter and FLOP reduction of the joint configurations is determined entirely by the filter width $K'$. The input mask zeros the non-important positions but does not remove the corresponding arithmetic, because the cross-channel mask selects scattered individual positions across the time--frequency grid rather than whole subcarriers or symbols. The dilated convolutions require a regular grid to operate on, so the masked positions must be retained as zeros to preserve that structure, and the dense first-layer convolution therefore still processes the full $52 \times 100$ input. The parameter and FLOP values consequently match the filter-width-only results in Table~\ref{tab:compression}. At $K' = 210$, the joint configuration retains 4\,392\,040 parameters (32.3\% reduction) and operates at 0.456~GFLOPs per inference frame, while the input representation has been reduced to 19.6\% of the original time--frequency grid. At $K' = 160$, the joint configuration retains 2\,578\,340 parameters (60.2\% reduction) and 0.267~GFLOPs.

\subsubsection{OBSERVATIONS}

The joint compression results suggest two practical observations. First, the input mask determines the achievable \ac{NMSE} floor in this configuration, and reducing the filter width below $K' = 160$ adds parameter savings without proportionate \ac{NMSE} cost relative to the cross-channel model. Second, the joint configuration at $K' = 160$ achieves a 60.2\% parameter reduction and a 60.4\% FLOP reduction while operating on 19.6\% of the time--frequency grid, which makes it a candidate operating point for deployment scenarios constrained on both compute and input pipeline cost. A joint optimisation that searches over input mask and filter width simultaneously, rather than applying the two reductions independently, is a direction we leave for future work.

\section{DISCUSSION}
\label{sec:discussion}

\subsection{TWO-LEVEL INTERPRETABILITY HIERARCHY}

The input-level and filter-level analyses reveal a consistent two-level interpretability structure. At the input, 80.4\% of features fall below the cross-channel relevance threshold and the cross-channel input correlation exceeds $r > 0.81$. At the internal filter level, 18.1\% of filters fall below the cross-channel relevance threshold, and the cross-channel filter correlation reaches $r = 0.9963$. The network progressively concentrates the information it relies on: input patterns are moderately shared across channel models, but the internal representations are near-identical across channel models (Fig.~\ref{fig:correlation}). 

The 23.8\% universal filter fraction lies in a similar range to the 19.6\% cross-channel input feature fraction, which suggests an approximate conservation of the essential information fraction across levels of abstraction. This two-level picture provides a richer account of the DPA-RDCNN's behaviour under multi-channel training than either level alone: the input-level result indicates \emph{which} time--frequency positions the network attends to, and the filter-level result indicates \emph{how} the resulting representations are organised internally.

\subsection{CROSS-CHANNEL GENERALISATION MECHANISM}

The $r = 0.9963$ filter correlation is the strongest empirical signal reported in this work. Multi-channel training does not appear to improve average performance across channel models simply by widening coverage. Instead, it converges to a shared internal filter representation, with the 58\% of filters classified as environment-specific modulating activation levels to accommodate channel-to-channel variation without altering the underlying feature space. The three \ac{OOD} channels engage the same internal representation, allowing the network to operate on propagation conditions it has not seen during training. The compression results of Section~\ref{sec:ood_pruning} are consistent with this account: at every compression level, \ac{OOD} \ac{NMSE} degrades approximately as fast as or slightly slower than \ac{ID} \ac{NMSE} (e.g., $0.57 \pm 0.15$\,\si{\decibel} vs. $0.59 \pm 0.48$\,\si{\decibel} at $K' = 160$; $1.90 \pm 0.15$\,\si{\decibel} vs. $2.21 \pm 0.20$\,\si{\decibel} at $K' = 96$), and the seed-to-seed variability on \ac{OOD} channels is consistently lower. The \ac{BER} curves of Section~\ref{sec:ber_pruning} reflect the same pattern qualitatively, and the \ac{NMSE} results confirm the model-specific differences directly.

\section{CONCLUSION}
\label{sec:conclusion}

This paper presented REACH, a relevance-based interpretability and compression framework for deep-learning vehicular channel estimators, and applied it to a multi-channel-trained DPA-RDCNN at both the input and internal filter levels. The two levels reveal a structural asymmetry: input relevance patterns are moderately shared across channel conditions (mean Pearson correlation $\approx 0.89$), while internal filter relevance is near-identical ($r = 0.9963$). Multi-channel training therefore converges to a shared internal representation rather than maintaining distinct subnetworks for different propagation conditions, and the three \ac{OOD} channels engage this representation with the same relevance ranking as the three \ac{ID} channels on which the network was trained. This provides a representational account of the cross-channel generalisation behaviour observed under multi-channel training. The filter taxonomy (23.8\% universal, 58.0\% environment-specific, 18.1\% redundant) is consistent across all five residual blocks, and block-level relevance varies by less than one percentage point. The network operates in a distributed mode in which every block contributes approximately equally, which means meaningful compression must operate at the filter level rather than the block level. Guided by this taxonomy, the compact models reduce both parameter count and FLOPs while preserving accuracy. At the practical operating point $K' = 210$, the reduction is 32.3\% with an average \ac{NMSE} change of $+0.16 \pm 0.45$\,\si{\decibel}. At $K' = 160$, the reduction reaches 60.2\% with an \ac{NMSE} change of $+0.58 \pm 0.36$\,\si{\decibel}. The \ac{ID}--\ac{OOD} gap remains small at every compression level. Two limitations bound these findings. The compact models are retrained from scratch rather than pruned surgically from the baseline. A natural extension is online pruning via group-sparse regularisation during training, which would allow the low-relevance filters identified by the taxonomy to converge towards small magnitudes and be removed surgically without retraining a separate compact model from scratch. The same attribution framework also extends to weight-level quantisation, where universal filters would be allocated higher bit budgets than redundant filters, providing a principled basis for non-uniform quantisation in deployed vehicular receivers. 
Finally, the analysis is conditioned on the synthetic Acosta-Marum channel models used throughout. The extent to which the observed representational structure and compression headroom transfer to measured vehicular channels, whose statistics may differ from the simulated models, is an open question for future work.

\bibliographystyle{IEEEtran}
\bibliography{ref}

@article{9810508,
  author={Hou, Jun and Liu, Huaijie and Zhang, Yang and Wang, Wei and Wang, Jiaqian},
  journal={IEEE Wireless Communications Letters},
  title={{GRU-Based Deep Learning Channel Estimation Scheme for the IEEE 802.11p Standard}},
  year={2023},
  volume={12},
  number={5},
  pages={764--768},
  doi={10.1109/LWC.2022.3187110}
}

@article{pan2021channel,
  title={{Channel Estimation Based on Deep Learning in Vehicle-to-Everything Environments}},
  author={Pan, Jing and Shan, Hangguan and Li, Rongpeng and Wu, Yingxiao and Wu, Weihua and Quek, Tony Q. S.},
  journal={IEEE Communications Letters},
  volume={25},
  number={6},
  pages={1891--1895},
  year={2021},
  publisher={IEEE}
}

@inproceedings{10621232,
  author={Gizzini, Abdul Karim and Medjahdi, Yahia and Mabrouk, Mouna Ben},
  booktitle={2024 IEEE International Mediterranean Conference on Communications and Networking (MeditCom)},
  title={{GRACE: Gradient-Based XAI Scheme for Channel Estimation in Wireless Communications}},
  year={2024},
  pages={572--577},
  doi={10.1109/MeditCom61057.2024.10621232}
}

@article{10368353,
  author={Gizzini, Abdul Karim and Medjahdi, Yahia and Ghandour, Ali J. and Clavier, Laurent},
  journal={IEEE Transactions on Vehicular Technology},
  title={{Towards Explainable AI for Channel Estimation in Wireless Communications}},
  year={2024},
  volume={73},
  number={5},
  pages={7389--7394},
  doi={10.1109/TVT.2023.3345632}
}

@article{bach2015pixel,
  title={{On Pixel-Wise Explanations for Non-Linear Classifier Decisions by Layer-Wise Relevance Propagation}},
  author={Bach, Sebastian and Binder, Alexander and Montavon, Gr{\'e}goire and Klauschen, Frederick and M{\"u}ller, Klaus-Robert and Samek, Wojciech},
  journal={PLOS ONE},
  volume={10},
  number={7},
  pages={e0130140},
  year={2015},
  publisher={Public Library of Science}
}

@article{acosta2007six,
  title={{Six Time- and Frequency-Selective Empirical Channel Models for Vehicular Wireless LANs}},
  author={Acosta-Marum, Guillermo and Ingram, Mary Ann},
  journal={IEEE Vehicular Technology Magazine},
  volume={2},
  number={4},
  pages={4--11},
  year={2007},
  publisher={IEEE}
}

@inproceedings{gizzini2021temporal,
  title={{Temporal Averaging LSTM-Based Channel Estimation Scheme for IEEE 802.11p Standard}},
  author={Gizzini, Abdul Karim and Chafii, Marwa and Ehsanfar, Shahab and Shubair, Raed M.},
  booktitle={2021 IEEE Global Communications Conference (GLOBECOM)},
  pages={01--07},
  year={2021},
  organization={IEEE}
}

@article{gizzini2020deep,
  title={{Deep Learning Based Channel Estimation Schemes for IEEE 802.11p Standard}},
  author={Gizzini, Abdul Karim and Chafii, Marwa and Nimr, Ahmad and Fettweis, Gerhard},
  journal={IEEE Access},
  volume={8},
  pages={113751--113765},
  year={2020},
  publisher={IEEE}
}

@article{10314524,
  author={Gizzini, Abdul Karim and Chafii, Marwa},
  journal={IEEE Transactions on Machine Learning in Communications and Networking},
  title={{RNN Based Channel Estimation in Doubly Selective Environments}},
  year={2024},
  volume={2},
  pages={1--18},
  doi={10.1109/TMLCN.2023.3332021}
}

@inproceedings{Ngorima2024,
  author={Ngorima, Simbarashe Aldrin and Helberg, Albert and Davel, Marelie H.},
  title={{A Data Pilot-Aided Temporal Convolutional Network for Channel Estimation in IEEE 802.11p Vehicle-to-Vehicle Communications}},
  booktitle={Southern Africa Telecommunication Networks and Applications Conference (SATNAC)},
  year={2024}
}

@misc{wandb,
  title={{Weights \& Biases: Developer Tools for ML}},
  author={Biewald, Lukas},
  year={2020},
  howpublished={\url{https://wandb.ai}},
  note={{Software available from wandb.com}}
}

@article{ye2018power,
  author={Ye, Hao and Li, Geoffrey Ye and Juang, Biing-Hwang},
  journal={IEEE Wireless Communications Letters},
  title={{Power of Deep Learning for Channel Estimation and Signal Detection in OFDM Systems}},
  year={2018},
  volume={7},
  number={1},
  pages={114--117},
  doi={10.1109/LWC.2017.2757490}
}

@article{soltani2019deep,
  title={{Deep Learning-Based Channel Estimation}},
  author={Soltani, Mehran and Pourahmadi, Vahid and Mirzaei, Ali and Sheikhzadeh, Hamid},
  journal={IEEE Communications Letters},
  volume={23},
  number={4},
  pages={652--655},
  year={2019},
  publisher={IEEE}
}

@inproceedings{li2017pruning,
  author={Li, Hao and Kadav, Asim and Durdanovic, Igor and Samet, Hanan and Graf, Hans Peter},
  title={{Pruning Filters for Efficient ConvNets}},
  booktitle={International Conference on Learning Representations (ICLR)},
  year={2017}
}

@inproceedings{he2018soft,
  author={He, Yang and Kang, Guoliang and Dong, Xuanyi and Fu, Yanwei and Yang, Yi},
  title={{Soft Filter Pruning for Accelerating Deep Convolutional Neural Networks}},
  booktitle={Proceedings of the Twenty-Seventh International Joint Conference on Artificial Intelligence (IJCAI)},
  pages={2234--2240},
  year={2018}
}

@article{yeom2021pruning,
  title={{Pruning by Explaining: A Novel Criterion for Deep Neural Network Pruning}},
  journal={Pattern Recognition},
  volume={115},
  pages={107899},
  year={2021},
  issn={0031-3203},
  doi={10.1016/j.patcog.2021.107899},
  author={Yeom, Seul-Ki and Seegerer, Philipp and Lapuschkin, Sebastian and Binder, Alexander and Wiedemann, Simon and M{\"u}ller, Klaus-Robert and Samek, Wojciech}
}

@article{he2024structured,
  author={He, Yang and Xiao, Lingao},
  title={{Structured Pruning for Deep Convolutional Neural Networks: A Survey}},
  journal={IEEE Transactions on Pattern Analysis and Machine Intelligence},
  volume={46},
  number={5},
  pages={2900--2919},
  year={2024},
  doi={10.1109/TPAMI.2023.3334614}
}

@article{cheng2024survey,
  author={Cheng, Hongrong and Zhang, Miao and Shi, Javen Qinfeng},
  title={{A Survey on Deep Neural Network Pruning: Taxonomy, Comparison, Analysis, and Recommendations}},
  journal={IEEE Transactions on Pattern Analysis and Machine Intelligence},
  volume={46},
  number={12},
  pages={10558--10578},
  year={2024},
  doi={10.1109/TPAMI.2024.3447085}
}

@article{yuan2024pruned,
  title={{Pruned Tree-Structured Temporal Convolutional Networks for Quality Variable Prediction of Industrial Process}},
  author={Yuan, Changqing and Xie, Yongfang and Xie, Shiwen and Wang, Jie},
  journal={Journal of Process Control},
  volume={143},
  pages={103312},
  year={2024},
  publisher={Elsevier}
}

@article{ngorima_access2026,
  author={Ngorima, S. A. and Helberg, A. S. J. and Davel, M. H.},
  title={{Convolutional DPA Refinement for Real-Time Channel Estimation in High-Mobility Vehicular Networks}},
  journal={IEEE Access},
  year={2026},
  note={{Under Review}}
}

@article{adadi2018peeking,
  title={{Peeking Inside the Black-Box: A Survey on Explainable Artificial Intelligence (XAI)}},
  author={Adadi, Amina and Berrada, Mohammed},
  journal={IEEE Access},
  volume={6},
  pages={52138--52160},
  year={2018},
  publisher={IEEE}
}

@inproceedings{simonyan2013deep,
  title={{Deep Inside Convolutional Networks: Visualising Image Classification Models and Saliency Maps}},
  author={Simonyan, Karen and Vedaldi, Andrea and Zisserman, Andrew},
  booktitle={Workshop at the 2nd International Conference on Learning Representations (ICLR)},
  year={2014}
}

@inproceedings{ancona2017unified,
  title={{Towards Better Understanding of Gradient-Based Attribution Methods for Deep Neural Networks}},
  author={Ancona, Marco and Ceolini, Enea and {\"O}ztireli, Cengiz and Gross, Markus},
  booktitle={Proceedings of the 6th International Conference on Learning Representations (ICLR)},
  year={2018}
}

@inproceedings{shrikumar2017learning,
  author    = {Shrikumar, Avanti and Greenside, Peyton and Kundaje, Anshul},
  title     = {{Learning Important Features Through Propagating Activation Differences}},
  booktitle = {Proceedings of the 34th International Conference on Machine Learning (ICML)},
  year      = {2017}
}

@ARTICLE{khan2026compression,
  author={Khan, Fazal Muhammad Ali and Hallaq, Mohammad and Abou-Zeid, Hatem and Erak, Omar and Waqar, Omer and Hassan, Syed Ali and Alhussein, Omar and Hossain, Ekram},
  journal={IEEE Communications Surveys \& Tutorials}, 
  title={{Model Compression for Sustainable AI in xG Wireless Networks: Recent Advances, Challenges, and Future Directions}}, 
  year={2026},
  volume={28},
  number={},
  pages={5747-5791},
  doi={10.1109/COMST.2026.3682638}}
\end{document}